\documentclass[prd,nofootinbib,superscriptaddress,preprint,floatfix]{revtex4-1}
\pdfoutput=1
\usepackage{dcolumn}
\usepackage{bm}
\usepackage{graphicx}
\usepackage{amssymb,amsmath,mathrsfs}
\usepackage{multirow}
\usepackage{afterpage}
\usepackage{bbm}
\usepackage[colorlinks=true
,urlcolor=blue
,anchorcolor=blue
,citecolor=blue
,filecolor=blue
,linkcolor=blue
,menucolor=blue
,linktocpage=true
,pdfproducer=medialab
,pdfa=true
]{hyperref}

\usepackage{tikz-feynman}
\tikzfeynmanset{compat=1.1.0}
\usetikzlibrary{trees}
\usetikzlibrary{decorations.pathmorphing}
\usetikzlibrary{decorations.markings}

\usepackage{array,booktabs}

\usepackage{slashed}
\usepackage{epsfig,psfrag,rotating,soul}
\usepackage{rotfloat}
\usepackage{empheq}  

\evensidemargin \oddsidemargin
\allowdisplaybreaks

\let\OLDthebibliography\thebibliography
\renewcommand\thebibliography[1]{
  \OLDthebibliography{#1}
  \setlength{\parskip}{0pt}
  \setlength{\itemsep}{0pt plus 0.3ex}
}

\newcommand{\beq}{\begin{equation}}
\newcommand{\eeq}{\end{equation}}

\renewcommand{\vec}{\bf}

\usepackage{xspace}

\newcommand{\ba}{\begin{array}}
	\newcommand{\ea}{\end{array}}
\newcommand{\bd}{\begin{displaymath}}
	\newcommand{\ed}{\end{displaymath}}
\newcommand{\besub}{\begin{subequations}}
	\newcommand{\eesub}{\end{subequations}}
\newcommand{\bea}{\begin{eqnarray}}
	\newcommand{\eea}{\end{eqnarray}}

\def\q2 {q^2}

\def\bt{\begin{table}}
	\def\et{\end{table}}

\tikzfeynmanset{every dot/.style={minimum size=5pt}}

\definecolor{mygray}{gray}{0.85} 
\definecolor{myblue}{cmyk}{0.65, 0.37, 0.0, 0.19}
\allowdisplaybreaks

\begin{document}

\title{Bounds on SMEFT affecting multi gauge and Higgs-gauge couplings using two and three body spin correlations in $e^-e^+\to 3l2j\slashed{E}$ process}

\author{Amir Subba}
\email{amirsubba@iitg.ac.in}
\affiliation{Department of Physics, Indian Institute of Technology Guwahati,\\ Guwahati, Assam 781039, India}
\affiliation{Department of Physical Sciences,\\ Indian Institute of Science Education and Research Kolkata, Mohanpur, West Bengal 741246, India}

\author{Ritesh K. Singh}
\email{ritesh.singh@iiserkol.ac.in}
\affiliation{Department of Physical Sciences,\\ Indian Institute of Science Education and Research Kolkata, Mohanpur, West Bengal 741246, India}

\begin{abstract}
	The upcoming electron-positron collider provides an ideal place to probe deviation from the Standard Model predictions with its clean environment, beam polarization and significant luminosity. We studied anomalous quartic gauge boson couplings~($VVW^-W^+, V\in\{\gamma,Z\}$), triple gauge couplings~($W^-W^+\gamma/Z)$, and Higgs-gauge couplings $(HVV, V\in\{W^\pm,Z,\gamma\}$) induced by $SU(2)_L \times U(1)_Y$ gauge invariant dimension-6 operators in $3l2j\slashed{E}$ final events with initial beam polarization. The phase space of two prominent amplitudes i.e, triple gauge boson production $(WWV)$ and vector boson scattering sub-processes, are selected with boosted decision trees. We employ the asymmetries related to polarization and spin correlation observables along with cross~section to constrain the anomalous couplings. The parity odd polarizations and spin correlations of jets from $W$ boson require flavor tagging which is done using artificial neural networks. We provide one parameter limits at $95\%$ confidence level combining cross~section and spin related observables. Finally, marginalized limits on all nine anomalous couplings are obtained with MCMC analysis. The limits are found to be insensitive to systematic errors and dominated by statistics. We find that the VBS-like events provides a tighter constraints to $c_W$ and $c_B$ Wilson coefficients (WCs) in comparison to limits from WWZ phase space. For the remaining $7$ WCs, WWZ contribute significantly to the overall limits.
\end{abstract}

\maketitle 

\section{Introduction}
	\label{sec:intro}
	Triple and Quartic gauge boson couplings are a direct consequence of non-abelian structure of $SU(3)_C\times SU(2)_L\times U(1)_Y$ gauge structure of the Standard Model~(SM) of particle physics. The precise measurement of these couplings would test the validity of the SM to a high accuracy. The structure of trilinear gauge couplings~(TGC) are probed via $VV,V\in\{W^\pm,\gamma,Z\}$ di-boson production process at various experiments like L3~\cite{L3:2004lwm,L3:1999znj,L3:1999cbg,L3:1998qun,L3:2004ulv}, ALEPH~\cite{ALEPH:1999jcv,ALEPH:1999ljd,ALEPH:2004klc}, DELPHI~\cite{DELPHI:1999pkm,DELPHI:1999ypr,DELPHI:1997vbx,DELPHI:1997ytm}, OPAL~\cite{OPAL:1997yzg,OPAL:1998ixj,OPAL:1998riq,OPAL:2003xqq}, D0~\cite{D0:2004fqq,D0:2006eed,D0:2009xgd}, CDF~\cite{CDF:1995vgr,CDF:2005xet,CDF:2007aqs}, ATLAS~\cite{ATLAS:2011vfj,ATLAS:2012mec,ATLAS:2012upi,ATLAS:2016bkj,ATLAS:2016nmw,ATLAS:2016zwm,ATLAS:2017bbg,ATLAS:2018mxa,ATLAS:2019rob}, and CMS~\cite{CMS:2013ant,CMS:2011egr,CMS:2015tmu,CMS:2013piy,CMS:2020mxy}. The quartic gauge boson couplings~(QGC) are probed experimentally via vector boson scattering~(VBS) in ATLAS~\cite{ATLAS:2014jzl,ATLAS:2016bkj,ATLAS:2018mxa,ATLAS:2019cbr,ATLAS:2019thr,ATLAS:2016nmw}, and CMS~\cite{CMS:2020gfh,CMS:2014mra,CMS:2017fhs,CMS:2019uys,CMS:2019qfk}. At the leading order in SM, the QGC could also be probed through the production of three massive gauge bosons ($V_1V_2V_3, V_i \in \{W,Z\}$). The CMS collaboration~\cite{CMS:2020hjs} at the LHC has observed such processes with 137 fb$^{-1}$ datasets. These observations were made in final states involving three, four, five, and six leptons, as well as in channels with two same-sign leptons accompanied by one or two jets. Similar searches for triboson production have been conducted by other experiments. For instance, the ATLAS collaboration analyzed final states containing $WWW, WWZ,$ and $WZZ$ events~\cite{ATLAS:2019dny}, while dedicated searches for $WWW$ production are reported in~\cite{ATLAS:2016jeu,CMS:2019mpq}. The experimental results from these studies show a remarkable agreement with the predictions of the Standard Model (SM).
	\\
	One of fundamental question related to the particle is its mass, and the standard consensus is that the mass of all particles are generated via electroweak symmetry breaking or Higgs mechanism~\cite{Englert:1964et,Higgs:1964pj,Guralnik:1964eu}. The mechanism suggests the presence of Higgs field with non-zero vacuum expectation value (VEV), $v$, which is responsible for symmetry breaking. The search for the quantum of Higgs field known as Higgs boson was finally settled with the discovery of scalar particle with mass around $125$~GeV at LHC~\cite{ATLAS:2012yve,CMS:2012qbp}. The properties of scalar particle~(Higgs boson) has been studied in great detail in various experiments~\cite{ATLAS:2013mma,Diglio:2014vpa,CMS:2012qwq,CMS:2013uhr,ATLAS:2024vxc,ATLAS:2023owm,ATLAS:2023yqk,CMS:2023gjz,CMS:2023sdw,ATLAS:2022tnm,CMS:2022wpo,CMS:2022urr,CMS:2022kdx,CMS:2022uhn,CMS:2022dwd} and the results are compatible with the prediction of the SM Higgs boson.\\
	With the discovery of Higgs boson, SM has been the best experimentally validated theory at fundamental scale, yet there are many experimental and theoretical potholes which cannot be filled within the framework of the SM. The mass of the Higgs boson in presence of higher quantum corrections are unbounded from above, the enormous gap between the electroweak and the Planck scale, mass of neutrinos, structure of dark matter are few issues which SM completely fails to address. There are large number of literature which tries to explain the shortcoming of SM by including new symmetry, particles, and dimensions. The only shortcomings to these majority of models is the null result from the experiments till date. The absence of any clear evidence of new particles at energies up to several times the Higgs boson mass allows one to parameterize the effects of arbitrary new physics residing at energies $E \gg v$ on physical observables at the electroweak scale in terms of higher dimension operators. This formalism of expanding SM in a model independent way is commonly referred as effective field theory~(EFT). In this formalism, SM is expanded by adding a set of higher dimensional operators to its usual dimension-4 Lagrangian. The framework is constructed with a understanding that if the new physics is heavy, it can be integrated out while keeping its effects in the Wilson coefficients~(WCs). The higher dimensional operators are suppressed by some characteristic energy scale $\Lambda > \sqrt{s}_{\text{LHC}}$. In presence of such higher dimensional operators, the effective Lagrangian would be~\cite{Buchmuller:1985jz},
	\begin{equation}
		\label{eqn:efflag}
		\mathcal{L}_{\text{Eff}} = \mathcal{L}_{SM} + \sum_{d>4}\sum_i\frac{c_i^{(d)}}{\Lambda^{d-4}}\mathscr{O}_i^{(d)},
	\end{equation}
	where $c_i^{(d)}$ are the WCs of $d$-dimensional operators $\mathscr{O}_i^{(d)}$. One can expect that $\Lambda$ to be much larger than the electroweak scale, thus the higher order terms of Eq.~(\ref{eqn:efflag}) are largely suppressed allowing one to truncate the expansion at some order of choice. However, this naive truncation based on dimensionality should be done with utmost care. In some of the di-boson processes, the interference between the SM and dim-6 operators remains substantially suppressed~\cite{Azatov:2016sqh}. However, it has been shown that these interference can be resurrected with final state with multi particles or with correlations which we have considered in our current work.\\
	We consider a set of dim-6 effective operators that induce changes in triple gauge boson couplings $(W^-W^+\gamma/Z)$, Higgs couplings with gauge bosons $(HVV, V \in \{W^\pm, Z, \gamma\})$, and quartic gauge boson couplings $(VV^\prime W^-W^+, V, V^\prime \in \{\gamma, Z\})$. We investigate these operators in final states with three leptons~$(3l)$, two jets~(2j), and missing energy~$(\slashed{E})$ at $e^-e^+$ collider with a center-of-mass energy of $1$ TeV, using initially polarized beams.\\
	The process includes a set of amplitudes that allow simultaneous study of the various anomalous couplings described above within a gauge-invariant framework. The desired final states can result from vector boson scattering and triple gauge boson production amplitudes. We select the phase space representing these two dominant amplitudes using boosted decision trees.\\
	To analyze the polarizations and spin correlations of the weak gauge bosons, we use the angular distribution of their decay products. For hadronic decays, jet flavors are reconstructed using an artificial neural network, as flavor tagging is necessary to prevent parity odd asymmetries from averaging out. For leptonic decays, we assume ideal identification of the final state leptons. These spin-related observables, along with the cross-section, are used to constrain the Wilson coefficients of higher-dimensional operators. We analyze the limits under different luminosities and systematic errors using Monte Carlo Markov chain techniques. \\
	The rest of the paper is structured as follows; in section~\ref{sec:anom} we describe the anomalous $W^-W^+\gamma/Z$, $HVV,V\in\{W^\pm,Z,\gamma\}$, $W^-W^+\gamma Z$, and $W^-W^+ZZ$ couplings in terms of nine independent dimension-6 operators. In section~\ref{sec:spin}, we describe event selection with boosted decision trees, flavor tagging using artificial neural network and the construction of spin observables. In section~\ref{sec:probe}, we describe the probe of anomalous couplings and the limits obtained. We finally conclude in section~\ref{sec:conclude}.

	\section{Operators in the electroweak sector}
	\label{sec:anom}
	Since the discovery of Higgs boson at LHC, the electroweak symmetry breaking is considered to be broken linearly and within this premise there are number of higher dimensional operators that could induce non-standard effects to TGC, QGC and HVV couplings. In this article, we consider the lowest order~($d=6$) operators\footnote{There is one more operator in the bosonic sector, $\mathscr{O}_{DW} =\text{Tr}\left([D_\mu,W_{\nu\rho}][D^\mu,W^{\nu\rho}]\right)$ but it is usually traded using equation of motion for $\mathscr{O}_{WWW}$ and the fermionic operators~\cite{Corbett:2015cfe} which we do not consider in the current work. Thus, it is neglected here.} assuming the lepton and baryon number conservation. The list of dim-6 operators in the bosonic sector affecting TGC, QGC and HVV in the Hagiwara-Ishihara-Szalapski-Zeppenfeld (HISZ) basis are~\cite{Degrande:2012wf,Degrande:2013rea,Hagiwara:1993ck}
\begin{equation}
\boxed{
\begin{aligned}
\mathscr{O}_{WWW} &= \text{Tr}[W_{\mu\nu}W^\mu_\rho W^{\nu\rho}], &
\mathscr{O}_{W} &= (D_\mu\Phi)^\dagger W^{\mu\nu} (D_\nu\Phi), &
\mathscr{O}_B &= (D_\mu \Phi)^\dagger B^{\mu\nu} (D_\nu \Phi), \\[0.5em]
\mathscr{O}_{WW} &= \Phi^\dagger W_{\mu\nu}W^{\mu\nu} \Phi, &
\mathscr{O}_{BB} &= \Phi^\dagger B_{\mu\nu}B^{\mu\nu}\Phi, &
\mathscr{O}_{BW} &= \Phi^\dagger B_{\mu\nu}W^{\mu\nu}\Phi, \\[0.5em]
\mathscr{O}_{\widetilde{W}WW} &= \text{Tr}[\widetilde{W}_{\mu\nu}W^\mu_\rho W^{\nu\rho}], &
\mathscr{O}_{\widetilde{W}} &= (D_\mu\Phi)^\dagger \widetilde{W}^{\mu\nu}(D_\nu\Phi), &
\mathscr{O}_{\widetilde{B}} &= (D_\mu\Phi)^\dagger \widetilde{B}^{\mu\nu}(D_\nu\Phi), \\[0.5em]
\mathscr{O}_{\widetilde{W}W} &= \Phi^\dagger \widetilde{W}_{\mu\nu}W^{\mu\nu} \Phi, &
\mathscr{O}_{\widetilde{B}B} &= \Phi^\dagger \widetilde{B}_{\mu\nu}B^{\mu\nu}\Phi, &
\mathscr{O}_{B\widetilde{W}} &= \Phi^\dagger B_{\mu\nu}\widetilde{W}^{\mu\nu}\Phi
\end{aligned}
}
\label{eqn:dim6}
\end{equation}
Here, the covariant derivative and the field tensors are given as,
	\begin{equation*}
		\begin{aligned}
			&D_\mu &=&~~ \partial_\mu + i\frac{g_\text{w}}{2}\sigma^i W^i_\mu + i\frac{g_1}{2}B_\mu,\\
			& W_{\mu\nu} &=&~~ \frac{\sigma^I}{2}W_{\mu\nu}^I,\\
			&W_{\mu\nu}^I &=&~~ i g_{W}\left(\partial_\mu W^I_\nu - \partial_\nu W^I_\mu + g_W\epsilon^{IJK}W^J_\mu W^K_\nu\right),\\
			&B_{\mu\nu} &=&~~ i\frac{g_1}{2}(\partial_\mu B_\nu - \partial_\nu B_\mu),
		\end{aligned}
	\end{equation*}
	where $g_1,$ and $g_{W}$ are the gauge couplings and $\sigma$ are the Pauli matrices. The dual of the field tensor is defined as $\widetilde{V}_{\mu\nu} = \frac{1}{2}\epsilon_{\mu\nu\rho\sigma}V^{\rho\sigma}$ where Levi-Civita tensor follows the standard convention, $\epsilon_{0123}=1$. In Eq.~(\ref{eqn:dim6}) the $12$ operators can be classified in terms of its $CP$ structure as,
	\begin{equation}
		\label{eqn:wilson}
		\begin{aligned}
			\mathscr{O}_i^{E} &= \{\mathscr{O}_B,\mathscr{O}_W,\mathscr{O}_{WWW},\mathscr{O}_{W W},\mathscr{O}_{BB},\mathscr{O}_{BW}\},\\
			\mathscr{O}_i^{O} &= \{\mathscr{O}_{\widetilde{W}WW},\mathscr{O}_{\widetilde{W}},\mathscr{O}_{\widetilde{W}W},\mathscr{O}_{\widetilde{B}B},\mathscr{O}_{B\widetilde{W}},\mathscr{O}_{\widetilde{B}}\},
		\end{aligned}
	\end{equation}
	where $\mathscr{O}_i^E$ are the $CP$-even and $\mathscr{O}_i^O$ denotes the $CP$-odd operators, respectively. Out of the $6$ CP-odd operators, only four of them are linearly independent~\cite{Degrande:2013rea}. For example $\mathscr{O}_{\widetilde{B}}$, and $\mathscr{O}_{B\widetilde{W}}$ can be written as,
	\begin{equation}
		\begin{aligned}
			&\mathscr{O}_{\widetilde{B}} &=&\quad \mathscr{O}_{\widetilde{W}} + \frac{1}{2}\mathscr{O}_{\widetilde{W}W} - \frac{1}{2}\mathscr{O}_{\widetilde{B}B},\\
			&\mathscr{O}_{B\widetilde{W}} &=& \quad-2 \mathscr{O}_{\widetilde{W}} - \mathscr{O}_{\widetilde{W}W}.
		\end{aligned}
	\end{equation}
	The operators $\mathscr{O}_{WW}$ and $\mathscr{O}_{BB}$ apart from inducing anomalous contribution to vertex would also induce two point function of weak states of the form,
	\begin{equation}
		\begin{aligned}
			\mathscr{O}_{WW} &= -\frac{c_{WW}}{4\Lambda^2}g^2_W v^2 \left[W^+_{\mu\nu}W^{-\mu\nu} + \sin^2\theta_W (\partial_\mu A_\nu)^2 + \cos^2\theta_W (\partial_\mu Z_\nu)^2 - \frac{\sin(2\theta_W)}{2}A_{\mu\nu}Z^{\mu\nu}\right.\\&-\left.\frac{\sin^2\theta_W}{2}(\partial_\nu A_\mu)(\partial^\mu A^\nu)-\frac{\cos^2\theta_W}{2}(\partial_\nu Z_\mu)(\partial^\mu Z^\nu)+ \dots \right],\\
			\mathscr{O}_{BB} &= -\frac{c_{BB}}{4\Lambda^2}g_1^2 v^2 \left[\cos^2\theta_W (\partial_\mu A_\nu)^2 + \sin^2\theta_W (\partial_\mu Z_\nu )^2 + \frac{\sin(2\theta_W)}{2}A_{\mu\nu}Z^{\mu\nu} \right.\\&-\left.\cos^2\theta_W (\partial_\nu A_\mu)(\partial^\mu A^\nu) - \sin^2\theta_W (\partial_\nu Z_\mu )(\partial^\mu Z^\nu) + \dots \right],
		\end{aligned}
	\end{equation}
	where $\dots$ denotes the higher order Lorentz structures containing combination of three or higher number of weak and Higgs states. The parameters $c_{WW}$, and $c_{BB}$ are the associated WC of operators $\mathscr{O}_{WW}$, and $\mathscr{O}_{BB}$, respectively. Thus, operators $\mathscr{O}_{WW}$, and $\mathscr{O}_{BB}$ induces change in kinetic term of weak boson after symmetry breaking, thus these two requires the renormalization of the gauge fields and the couplings. However, one can redefined these operators as,
	\begin{equation}
		\begin{aligned}
			\mathscr{O}_{WW} &= (\Phi^\dagger\Phi - v^2/2)W_{\mu\nu} W^{\mu\nu},\\
			\mathscr{O}_{BB} &= (\Phi^\dagger\Phi - v^2/2)B_{\mu\nu}B^{\mu\nu}
		\end{aligned}
	\end{equation}
	to remove any residual part proportional to $v^2$, and this redefinition leads to the absence of any two point function along with TGC~\cite{Degrande:2013rea} in these operators. After the operator redefinition, only Higgs-gauge couplings exists at the three point level.\\
	The operator $\mathscr{O}_{BW}$ affects the $Z\gamma$ mixing, which leads to the shift in the mass eigenstates from those in the SM as
	\begin{equation}
    \begin{aligned}
        A_\mu &= \left(1 - \frac{g_1^2g_W^2v^2}{4(g_1^2+g_W^2)}\frac{c_{BW}}{\Lambda^2} \right) A_\mu^{\mathrm{SM}} + \frac{g_1g_Wv^2}{8} \frac{g_1^2-g_W^2}{g_1^2+g_W^2}\frac{c_{BW}}{\Lambda^2}Z_\mu^{\mathrm{SM}}\\
        Z_\mu &= \left(1+\frac{g_1^2g_W^2v^2}{4(g_1^2+g_W^2)}\frac{c_{BW}}{\Lambda^2}\right)Z_\mu^{\mathrm{SM}} + \frac{g_1g_Wv^2}{8} \frac{g_1^2-g_W^2}{g_1^2+g_W^2} \frac{c_{BW}}{\Lambda^2}A_\mu^{\mathrm{SM}}
    \end{aligned}
\end{equation}
	where $c_{BW}$ is the WC of $\mathscr{O}_{BW}$. The corresponding mass of $Z$ boson is,
	\begin{equation}
		m_Z^2 = \frac{g_{W}^2+g_1^2}{4}v^2\left[1-\frac{g_{W}^2g_1^2}{g_{W}^2+g_1^2}\frac{v^2}{\Lambda^2}c_{BW}\right].
	\end{equation}
	Thus, $\mathscr{O}_{BW}$ induces change in the $\rho$ parameter at the tree level, violating the custodial symmetry. We ignore this operators considering the precision of electroweak precision observables. We describe the relevant gauge and Higgs-gauge couplings below.\\ 	
	\textbf{Triple Gauge Boson Couplings:}
	The anomalous $W^-W^+\gamma/Z$ couplings can be parameterized in terms of $14$ parameters as~\cite{Hagiwara:1986vm}
	\begin{equation}
		\begin{aligned}
			\mathcal{L}_{WWV} &=i g_{WWV} (g^V_1(W_{\mu \nu}^+ W^{-\mu}-W^{+\mu}W^{-}_{\mu \nu})V^{\nu}+i g_4^V W^{+}_{\mu} W^{-}_{\nu}(\partial^{\mu} V^{\nu}+\partial^{\nu} V^{\mu})\\&-ig^V_5\epsilon^{\mu \nu \rho \sigma}(W_{\mu}^+ \partial_{\rho} W_{\nu}^- - \partial_{\rho} W_{\mu}^+ W_{\nu}^-)V_{\sigma}+\frac{\lambda^V}{m_W^2}W_{\mu}^{+\nu} W_{\nu}^{-\rho} V_{\rho}^{\mu}+\frac{\lambda^{\tilde{V}}}{m_W^2}W_{\mu}^{+\nu} W_{\nu}^{-\rho} \tilde{V}_{\rho}^{\mu}\\&+\kappa^V W^{+}_{\mu} W_{\nu}^{-} V^{\mu \nu}
			+\kappa^{\tilde{V}} W^{+}_{\mu} W_{\nu}^{-} \tilde{V}^{\mu \nu}).
		\end{aligned}
	\end{equation} 
	The couplings $g^V_1$, $\lambda^V$ and $\kappa^V$ are $CP$-even, while $\lambda^{\tilde{V}}$ and $\kappa^{\tilde{V}}$ are $CP$-odd. Within the SM, $g_1^\gamma = g_1^Z=1$ and all other couplings are explicitly zero. The conservation of charge of $W$ boson implies that any contribution to $g_1^\gamma$ should vanish. The subset of operators listed in Eq.~(\ref{eqn:dim6}) affecting $W^-W^+\gamma/Z$ couplings are
	$$\mathscr{O}_i^{WWV} \in \{\mathscr{O}_B,\mathscr{O}_W,\mathscr{O}_{\widetilde{W}},\mathscr{O}_{WWW},\mathscr{O}_{\widetilde{W}WW}\}.$$ The anomalous part of the above $WWV$ Lagrangian can be expressed in terms of Wilson coefficient $(c_i)$ of dim-6 operators as~\cite{Degrande:2012wf}
	\begin{equation}
		\label{eqn:relation}
		\begin{aligned}
			&\Delta g_1^Z = c_W \frac{m_Z^2}{2\Lambda^2},\\
			& \Delta \kappa_\gamma = (c_W + c_B)\frac{m_W^2}{2\Lambda^2},\\
			& \Delta \kappa_Z = (c_W - c_B \tan^2\theta_W)\frac{m_W^2}{2\Lambda^2},\\
			&\lambda_\gamma = \lambda_Z = c_{WWW}\frac{3g_W^2m_W^2}{2\Lambda^2},\\
			& g_4^V = g_5^V = 0,\\
			&\widetilde{\kappa} = c_{\widetilde{W}}\frac{m_W^2}{2\Lambda^2},\\
		    & \widetilde{\kappa}_Z = - c_{\widetilde{W}}\tan^2\theta_W \frac{m_W^2}{2\Lambda^2},\\
		    &\widetilde{\lambda}_\gamma = \widetilde{\lambda}_Z = c_{\widetilde{W}WW}\frac{3g_W^2m_W^2}{2\Lambda^2}.
		\end{aligned}
	\end{equation}
	From the above Eq.~(\ref{eqn:relation}), one can obtain two more relations
	\begin{equation}
		\begin{aligned}
			\Delta g_1^Z &= \Delta \kappa_Z + \tan^2\theta_W \Delta \kappa_\gamma,\\
			0 &= \widetilde{\kappa}_Z + \tan^2\theta_W \widetilde{\kappa}_Z,
		\end{aligned}
	\end{equation}
	which reduces the independent anomalous couplings to five.
	 Many studies~\cite{Subba:2023rpm,Choudhury:2022iqz,CMS:2021icx,ATLAS:2021jgw,CMS:2021foa,Biekotter:2021int,Calfayan:2020tuk,Rahaman:2019lab,Koksal:2019oqt,Gutierrez-Rodriguez:2019hek,Billur:2019cav,Rahaman:2019mnz,CMS:2019ppl,CMS:2019efc,daSilvaAlmeida:2018iqo,Bellan:2018xxs,Cuevas:2018jah,CMS:2018hlo,Li:2017kfk,Burger:2017goy,ATLAS:2017pbb,CMS:2017egm,ATLAS:2017pbb,CMS:2017egm,ATLAS:2017luz,Iliadis:2017bqd,CMS:2016gct,Hassani:2016jnz,Becker:2016fju,Wang:2016zbh,Zhang:2016zsp,Falkowski:2016cxu,CMS:2016qth,ATLAS:2016qzn,Etesami:2016rwu,ATLAS:2016bkj,ATLAS:2016zwm,Falkowski:2015jaa,ATLAS:2014ofc,Biswal:2014oaa,CMS:2013ryd,CMS:2013ant,Han:2013jmw} have probed anomalous $W^+W^-\gamma/Z$ couplings employing various kinematic observables like differential rate and angular functions at both hadron and lepton collider.\\
	 \textbf{Higgs couplings to Gauge bosons:}
	 The operators inducing anomalous couplings of Higgs with gauge bosons are,
	 $$\mathscr{O}_i^{HVV}\in \{\mathscr{O}_B,\mathscr{O}_W,\mathscr{O}_{\widetilde{W}},\mathscr{O}_{WW},\mathscr{O}_{\widetilde{W}W},\mathscr{O}_{BB},\mathscr{O}_{\widetilde{B}B}\}.$$
	 These operators gives rise to Higgs interactions with gauge boson pairs that take the following form in the unitary gauge~\cite{Corbett:2012ja},
	 	\begin{equation}
	 		\begin{aligned}
	 			\mathscr{L}_{HVV} &= g_{H\gamma\gamma}HA_{\mu\nu}A^{\mu\nu} +  \widetilde{g}_{H\gamma\gamma}H\widetilde{A}_{\mu\nu}A^{\mu\nu}+ g_{HZ\gamma}^{(1)}A_{\mu\nu}Z^\mu\partial^\nu H + g_{HZ\gamma}^{(2)}HA_{\mu\nu}Z^{\mu\nu}  + \widetilde{g}_{HZ\gamma}^{(1)}\widetilde{A}_{\mu\nu}Z^\mu\partial^\nu H \\&+ \widetilde{g}_{HZ\gamma}^{(2)}H\widetilde{A}_{\mu\nu}Z^{\mu\nu} +  g_{HZZ}^{(1)}Z_{\mu\nu}Z^\mu\partial^\nu H + g_{HZZ}^{(2)}HZ_{\mu\nu}Z^{\mu\nu} + \widetilde{g}_{HZZ}^{(1)} \widetilde{Z}_{\mu\nu}Z^\mu \partial^\nu H + \widetilde{g}_{HZZ}^{(2)}H\widetilde{Z}_{\mu\nu} Z^{\mu\nu} \\&+   g_{HWW}^{(1)}(W^+_{\mu\nu}W^{-\mu}\partial^\nu H + H.c) + g_{HWW}^{(2)}HW^+_{\mu\nu} W^{-\mu\nu} + \widetilde{g}_{HWW}^{(1)}(\widetilde{W}^+_{\mu\nu}W^{-\mu}\partial^\nu H + H.c) \\&+ \widetilde{g}_{HWW}^{(2)}H\widetilde{W}^+_{\mu\nu}W^{-\mu\nu} .
	 		\end{aligned}
	 	\end{equation}
	 The effective couplings are related to the Wilson coefficient as
	 	\begin{align}
	 		g_{H\gamma\gamma} &= -\left(\frac{g^2_Wv\sin^2\theta_W}{\Lambda^2}\right)(c_{WW}+c_{BB}), \nonumber \\
	 		g_{HZ\gamma}^{(1)} &= \left(\frac{g^2_Wv\sin\theta_W}{4\cos\theta_W\Lambda^2}\right)(c_W-c_B), \nonumber \\
	 		g_{HZ\gamma}^{(2)} &= -\left(\frac{g^2_Wv\sin\theta_W\cos\theta_W}{\Lambda^2}\right)(c_{WW}-\tan^2\theta_W c_{BB}),\nonumber \\
	 		\widetilde{g}_{H\gamma\gamma} &= \left(\frac{g^2_Wv\sin^2\theta_W}{2\Lambda^2}\right)(c_{\widetilde{W}W} +c_{\widetilde{B}B}),\nonumber \\
	 		\widetilde{g}_{HZ\gamma}^{(1)} &= \left(\frac{g^2_Wv\sin\theta_W}{4\Lambda^2\cos\theta_W}\right)c_{\widetilde{W}},\nonumber \\
	 		\widetilde{g}_{HZ\gamma}^{(2)} &= \left(\frac{g^2_Wv\sin\theta_W\cos\theta_W}{2\Lambda^2}\right)(c_{\widetilde{W}W} - \tan^2\theta_Wc_{\widetilde{B}B}),\nonumber \\
	 		g_{HZZ}^{(1)} &= \left(\frac{g^2_Wv}{2\Lambda^2}\right)\frac{\cos^2\theta_Wc_W+\sin^2\theta_Wc_B}{2\cos^2\theta_W},\nonumber \\
	 		g_{HZZ}^{(2)} &= -\left(\frac{g^2_Wv}{2\Lambda^2}\right)\frac{\sin^2\theta_Wc_{BB}+\cos^4\theta_Wc_{WW}}{2\cos^2\theta_W},\nonumber \\
	 		\widetilde{g}_{HZZ}^{(1)} &= \left(\frac{g^2_Wv}{16\Lambda^2}\right)c_{\widetilde{W}},\nonumber \\
	 		\widetilde{g}_{HZZ}^{(2} &= -\frac{g^2_Wv\sin^2\theta_W\tan^2\theta_W}{4\Lambda^2}c_{\widetilde{W}W} + \frac{g^2_Wv\cos^2\theta_W}{8\Lambda^2}c_{\widetilde{B}B},\nonumber \\
	 		g_{HWW}^{(1)} &= \left(\frac{g^2_Wv}{2\Lambda^2}\right)\frac{c_W}{2},\nonumber \\
	 		g_{HWW}^{(2)} &= -\left(\frac{g_W^2v}{2\Lambda^2}\right)c_{WW},\nonumber \\
	 		\widetilde{g}_{HWW}^{(1} &= \frac{g^2_Wv}{16\Lambda^2\cos\theta_W\sin^2\theta_W}c_{\widetilde{W}},\nonumber \\
	 		\widetilde{g}_{HWW}^{(2)}&= -\frac{g^2_Wv}{4\Lambda^2}c_{\widetilde{W}W}
	 	\end{align}
	 The probe of Higgs couplings with gauge boson are of paramount importance to settle the discourse of $CP$ structure of Higgs, which till current time is considered to be even.  The anomalous $HVV$ couplings have been probed in studies~\cite{Hernandez-Juarez:2024zpk,Kniehl:1990mq,Soni:1993jc,Hernandez-Juarez:2023dor,Biswal:2005fh,Dutta:2008bh,Rao:2020hel,Godbole:2007cn,Cakir:2013bxa,Sahin:2019wew,Gauld:2023gtb,Gounaris:2000tb,He:2019kgh,Buchalla:2013mpa,Lagouri:2024ozw,Spor:2023sdk,DAgnolo:2023rnh,Fabbrichesi:2023jep,Lu:2022jzs,Sharma:2022epc,Kumar:2019bmk,Fabbrichesi:2023jep,Gabrielli:2013era,Gonzalez-Lopez:2020lpd,Asteriadis:2022ebf,Senol:2012fc,Buchalla:2015wfa,Biswal:2012mp,Corbett:2012dm,Corbett:2012ja,Davis:2021tiv,Shi:2018lqf,L3:2004vpt,Gonzalez-Garcia:1999ije} and references therein. The Higgs coupling involving $Z$ and photon, i.e. $HZ\gamma,H\gamma\gamma$ are interesting structure as they do not exist at SM tree level and arise at one loop level. CMS and ATLAS Collaborations~\cite{ATLAS:2023yqk} reports the first evidence for Higgs decay to $Z$ boson and photon with a statistical significance of $3.4$ standard deviation. Thus, it presents a strong area of interest to probe for new physics. One of the channel that predicts the existence of scalar with mass around $125$~GeV, was two photon channel~\cite{ATLAS:2012yve,CMS:2012qbp}. This channel would be a clean channel to look for any deviation from SM in upcoming $e^-e^+$ collider.
      \begin{figure*}[!htb]
	 	\centering
	 	\includegraphics[width=0.32\textwidth,height=4cm]{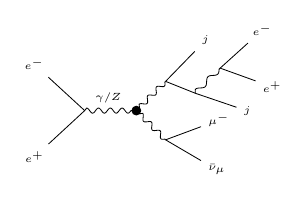}
	 	\includegraphics[width=0.32\textwidth,height=4cm]{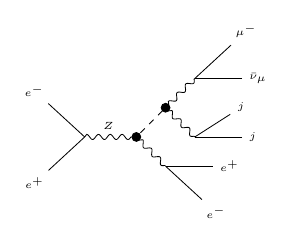}
	 	\includegraphics[width=0.32\textwidth,height=4cm]{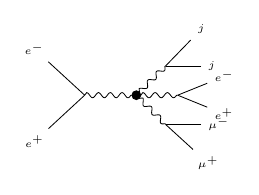}
	 	\includegraphics[width=0.32\textwidth,height=4cm]{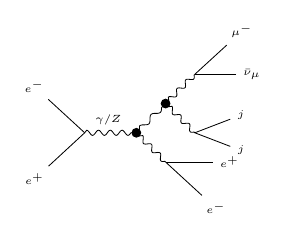}
	 	\includegraphics[width=0.32\textwidth,height=4cm]{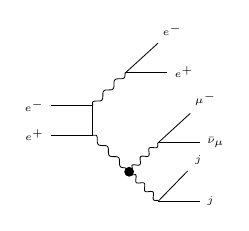}
	 	\includegraphics[width=0.32\textwidth,height=4cm]{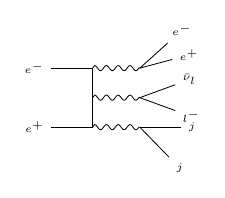}
	 	\includegraphics[width=0.32\textwidth,height=4cm]{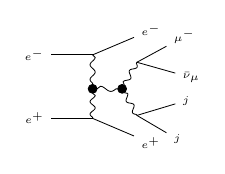}
	 	\includegraphics[width=0.32\textwidth,height=4cm]{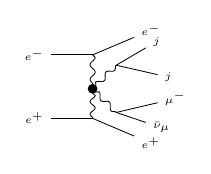}
	 	\includegraphics[width=0.32\textwidth,height=4cm]{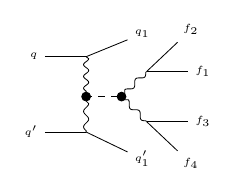}
	 	\caption{Representative gauge invariant set of Feynman diagrams for the electroweak production of three gauge bosons~(top and middle row) and VBS amplitudes~(bottom row) that would present the same final states. SMEFT operators modifies the triple and quartic couplings.}
	 	\label{fig:feynmandiag}
	 \end{figure*}
     \\
	 \textbf{Quartic Gauge boson couplings:}
	 The operators listed in Eq.~(\ref{eqn:dim6}) apart from affecting TGC and HVV could also directly induce anomalous behavior in quartic gauge boson couplings (QGC). The operators are,
	 $$\mathscr{O}_i^{WWVV} \in \{\mathscr{O}_W,\mathscr{O}_{WWW},\mathscr{O}_{WW},\mathscr{O}_{\widetilde{W}WW}\}.$$ The processes like vector boson scattering and tri-boson production are sensitive to the modifications of quartic couplings. These processes also depends on the tri-linear $W^-W^+\gamma/Z$, and $HVV$ couplings. From the operators point of view, a set of operators might affect different structures. For example, a set of operator affecting quartic couplings also induce change in triple gauge couplings. Thus, it becomes important from the sensitivity point of view that such set of operators must be studied in totality of all affected structures. Usually, each set of operators are studied in isolation. Di-boson process are mostly probed for $W^-W^+\gamma/Z$, and $HZZ$ couplings, nevertheless tri-boson processes becomes relevant comes from the helicity selection rules. Helicity selection rules suppress the linear, tree-level sensitivity of high energy transverse boson pair production to dimension-six SMEFT effects~\cite{Simmons:1989zs,Dixon:1993xd,Azatov:2016sqh}. The study of each set of operators in isolation becomes ambiguous due to each set affecting more than one set of couplings. Thus, a study of a process where all set of operators can come to life becomes very important to understand the marginalized constraint on them and also to gauge down the possibility of ultraviolet complete. To complement this ideas, we probed $2je^-e^+l^-\slashed{E}$ production process which simultaneously provides various amplitudes sensitive to all three sets of operators. A subset of schematic Feynman diagrams for $2je^-e^+\mu^-\bar{\nu}_\mu$ process at the leading order are shown in Fig.~\ref{fig:feynmandiag}. The dark blob represents the contribution from the dim-6 operators. The number of point of SMEFT insertion in the amplitudes lies between $0-2$, where zero insertion is the SM. \\
	 We implement the relevant operators in {\tt FeynRules}~\cite{Alloul:2013bka} to obtain {\tt Universal FeynRules Output}~(UFO)~\cite{Darme:2023jdn,Degrande:2011ua} model file. We used diagonal CKM matrix and``$\alpha$-scheme''~$\{M_Z,\alpha,G_F\}$ as an electroweak input scheme. The model file are used in Monte-Carlo generator {\tt MadGraph5$\_$aMC$@$NLO}~({\tt MG5} henceforth)~\cite{Alwall:2014hca} for event generation. 
	  \\
	  In the next section, we discuss the polarization and spin correlations in VBS and WWZ events and their reconstruction though asymmetries in angular functions. We also discuss the flavor tagging of final decayed jets from $W$ bosons using neural networks. 
 \section{spin asymmetries}
	  \label{sec:spin}
	  The upcoming lepton collider like FCC-ee~\cite{FCC:2018evy}, CLIC~\cite{CLICdp:2018cto} and ILC~\cite{ILC:2007bjz,ILC:2007oiw} would collide an initially polarized beams. The use of polarized beams would benefit the search of new physics~(NP) signal which are usually drowned in the sea of SM background. The cross~section in presence of longitudinal beam polarization behaves as~\cite{Moortgat-Pick:2005jsx},
	 \begin{equation*}
	 	\begin{aligned}
	 		\label{en:polxsec}
	 		\sigma(\eta_3,\xi_3) &= \frac{1}{4}\left[(\sigma_{\text{RL}}+\sigma_{\text{LR}})+(\eta_3-\xi_3)(\sigma_{\text{RL}}-\sigma_{\text{LR}})-\eta_3\xi_3(\sigma_{\text{RL}}+\sigma_{\text{LR}})\right],
	 	\end{aligned}	
	 \end{equation*}
	  where $\eta_3,\xi_3$ represents the degree of polarization for electron and positron beam, respectively. The $\sigma_{\text{RL}}(\sigma_{\text{LR}})$ is the cross~section when electron is $100\%$ right-handed~(left-handed) polarized and positron is $100\%$ left-handed~(right-handed) polarized. The equation suggest that on using the set of two flipped polarization would provide a directional cuts on differential distribution, which would help to constrain the parameters of new physics tighter.
	 We consider final states with three charged leptons, two jets and missing neutrino at $e^-e^+$ collider at $\sqrt{s}=1$~TeV in presence of initial beam polarization. At the leading order~(LO), some of the relevant Feynman diagrams resulting to the desired final state are shown in Fig.~\ref{fig:feynmandiag}. The process contains a sub-set of amplitudes that would corresponds to triple gauge boson production~(WWZ), and vector boson scattering~(VBS).
     \\
     We exploit two different phase space targeting two different sub-processes i.e, WWZ and VBS. We developed boosted decision trees~(BDT) to classify the events as WWZ and VBS-like events. To train a network we generated two set of events representing WWZ and VBS process. For WWZ process, we generate via. {\tt e-~ e+~>~w-~w+~z} command in {\tt MG5} and the VBS-like amplitudes are generated using diagram filter plugin in {\tt MG5} in which we remove WWZ-like diagrams.
     \begin{figure*}
	  	\centering
	  	\includegraphics[width=0.49\textwidth]{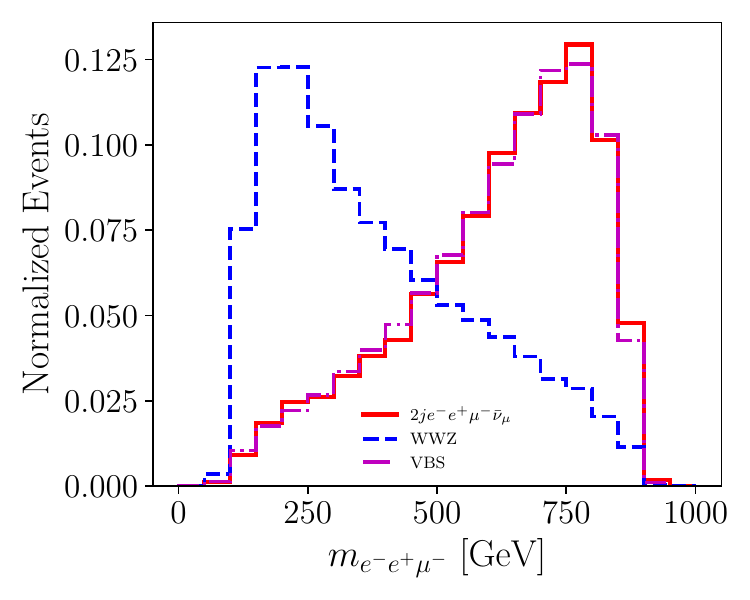}
	  	\includegraphics[width=0.49\textwidth]{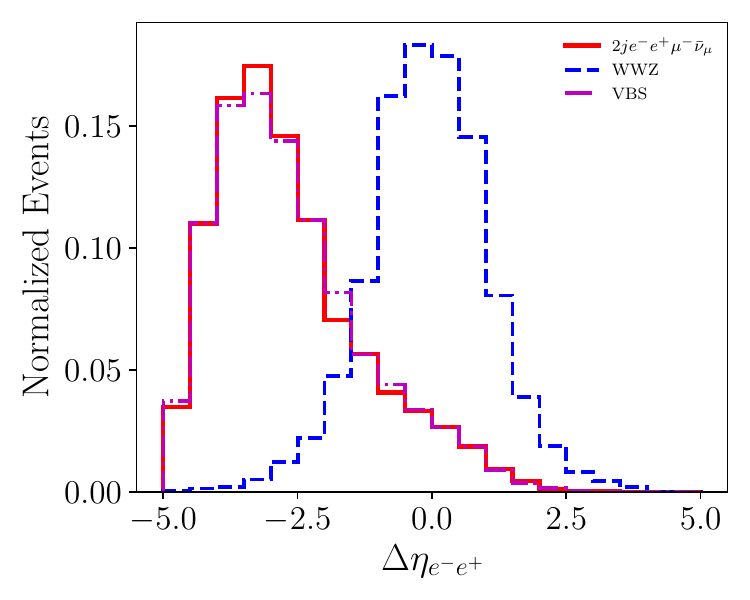}
	  	\caption{\label{fig:process}Parton level distribution of transverse momenta of muon (left panel), and pseudorapidity difference of electron and positron (right panel). The distribution are obtained for three different process, i) $e^-e^+\to 2j e^-e^+\mu^- \bar{\nu}_\mu$ (shown in solid red curve), ii) WWZ production $e^-e^+ \to W^+W^-Z$ (shown in dashed blue curve), and iii) VBS $e^-e^+ \to e^-e^+W^-W^+$ (depicted by dashed magenta curve) at $\sqrt{s}=1$ TeV. }
	  \end{figure*}
      The matrix level events are then passed to {\tt Pythia} for showering and hadronization. The final state hadrons are clustered using {\tt Anti-$k_T$} with jet radius $R=0.7$ followed by $k_T$~($R=1.0$) jet clustering algorithm available in {\tt Fastjet} package. In order to select the WWZ and VBS-like phase space, we develop boosted decision trees (BDT) to perform binary classification.
	  The features we used to train our BDT are invariant mass of electron-positron system~($m_{e^-e^+}$), and three leptons~($m_{3l})$, transverse momentum~($p_{T}^{jj},p_T^{ee},p_T^\mu,p_T^{3l},p_T^{e\mu},p_T^j$), pseudorapidity difference~($\Delta\eta_{jj},\Delta\eta_{ee},\Delta\eta_{e\mu}$), and transverse mass which is defined as $m_{T(\mu\nu)} = \sqrt{2E_{T\mu}\slashed{E}_T\left(1-\cos(\Delta\theta)\right)}$ with $\Delta\theta$ the angle between muon and reconstructed neutrino. We show the parton level distribution of some features in Fig~\ref{fig:process}. The model classify two events with an accuracy of $\approx 96\%$. Once the event classification is done, the two phase space are probed separately to obtain the polarizations and spin-correlations of weak boson described in spin density matrix.\\ 
       For a case when two spin-1 boson are co-produced, the density matrix can be written as,
	  \begin{equation}
	  	\label{eqn:spin1spin1}
	  	\begin{aligned}
	  		\rho_{(11)} = \frac{1}{9}\mathbb{I}\otimes \mathbb{I}+\sum_i p_i^{(1)}\mathbb{I}\otimes J_i + \sum_i p_i^{(2)} J_i \otimes \mathbb{I} + \sum_{i,j} p^{(12)}_{ij}J_i\otimes J_j,	
	  	\end{aligned}
	  \end{equation}
	  where $J_i$ (see Appendix~\ref{sec:ji}) are the matrices constructed from fundamental spin-1 operators, and $p_i^{(1/2)}$, $p_{ij}^{(12)}$ are the polarizations and spin-correlations parameters. Using the decay density matrix $\Gamma$~(see Appendix~\ref{sec:Nddm}), we can obtain the joint angular distribution form final decayed fermions of $W$ boson as,
      \begin{equation}
	  	\frac{1}{\sigma}\frac{d\sigma}{d\Omega^{(1)}d\Omega^{(2)}} = \sum \rho_{(11)} \times \left(\Gamma_{W^-}\otimes \Gamma_{W^+}\right).
	  \end{equation}
	  And the $80$ parameters of density matrix are obtained by partial integration of the joint angular distribution which are represented in terms of asymmetries of various angular functions. In general the asymmetries are computed numerically as,
	  \begin{equation}
	  	A_{ij}^{12} = \frac{\sigma(C_i^{(1)}C_j^{(2)} > 0) - \sigma(C_i^{(1)}C_j^{(2)} < 0)}{\sigma(C_i^{(1)}C_j^{(2)} > 0) + \sigma(C_i^{(1)}C_j^{(2)} < 0)},
	  \end{equation}
	  where the angular correlators $(C_i)$ are, 
      \begin{equation}
	  	\label{eqn:correlators}
	  	\begin{aligned}
	  		C_i \in &\left[C_1=1,C_2 = \cos\phi, C_3=\sin\phi,c_4=\cos\theta,C_5=C_2C_3,C_6=C_2C_4, C_7=C_3C_4,\right.\\&\left.C_8=C_2^2-C_3^2,C_9=\sqrt{C_4^2-1}(4C_4^2-1)\right].
	  	\end{aligned}    
	  \end{equation}
	  The polar~($\theta$) and azimuth~($\phi$) angle are obtained at the rest frame of $W$ boson; the beam line defines the $z$-axis in lab and the production plane of $W$ defines the $xz$ plane~($\phi=0$). Next we consider a process where three spin-1 particles are co-produced i.e., WWZ production process. Similarly the density matrix in the case of three spin-1 particle can be expanded as,
	  \begin{equation}
	  	\begin{aligned}
	  		\rho_{(111)} &= \frac{1}{27}\mathbb{I}_{27\times 27} + \sum_i p_i^{(1)} J_i\otimes\mathbb{I} \otimes \mathbb{I} + \sum_i p_i^{(2)}\mathbb{I}\otimes J_i \otimes \mathbb{I} + \sum_ip^{(3)}\mathbb{I}\otimes \mathbb{I}\otimes J_i\\&+ \sum_{i,j}p_{ij}^{(12)}J_i\otimes J_j\otimes \mathbb{I} + \sum_i p_{ij}^{(13)} J_i \otimes \mathbb{I}\otimes J_j +\sum_{ij} p_{ij}^{(23)}\mathbb{I}\otimes J_i \otimes J_j \\&+ \sum_{ijk} p^{(123)}_{ijk}J_i\otimes J_j\otimes J_k,
	  	\end{aligned}
	  \end{equation}
	  where $p^{(123)}$ represents the three particle spin-correlation parameters. Similarly to the two spin-1 particle case, the parameters for the three particle spin density are obtained in terms of asymmetries,
       \begin{equation}
	  	A_{ijk}^{123 } = \frac{\sigma(C_i^{(1)}C_j^{(2)}C_k^{(3)}>0) - \sigma(C_i^{(1)}C_j^{(2)}C_k^{(3)}<0)}{\sigma(C_i^{(1)}C_j^{(2)}C_k^{(3)}>0)+\sigma(C_i^{(1)}C_j^{(2)}C_k^{(3)}<0)},
	  \end{equation}
	  where the correlators $C_i$ are given in Eq.~(\ref{eqn:correlators}).
       \begin{figure*}[!htb]
	  	\centering
	  	\includegraphics[width=0.49\textwidth]{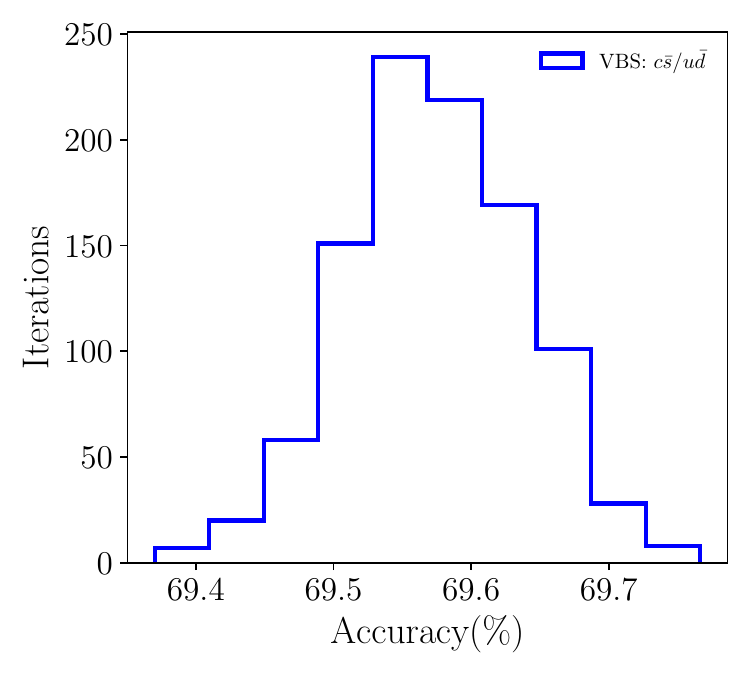}
	  	\includegraphics[width=0.49\textwidth]{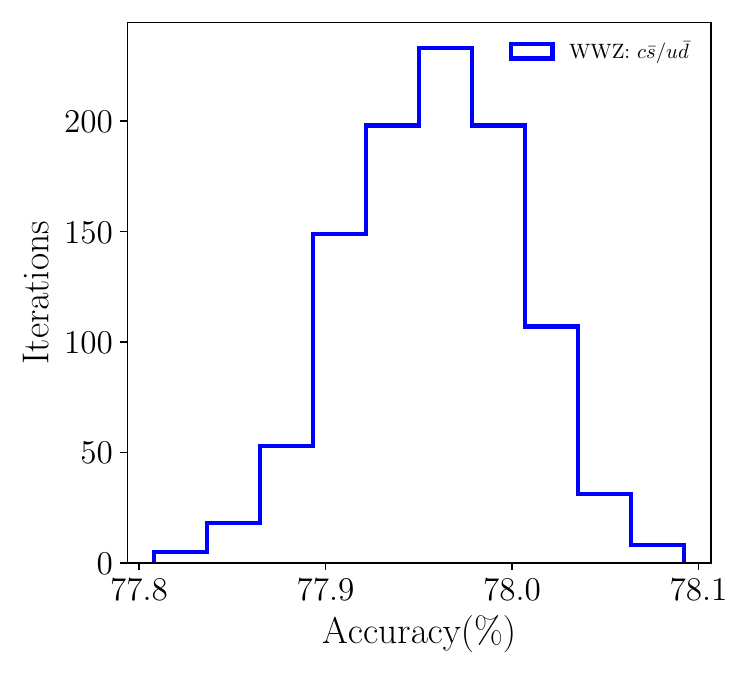}
	  	\caption{\label{fig:flavor}Distribution of accuracy on classifying two hardest jets as \emph{up/down} type jets using neural networks for VBS and WWZ events.}
	  \end{figure*}
      \begin{figure*}[!htb]
	  	\centering
	  	\includegraphics[width=0.98\linewidth]{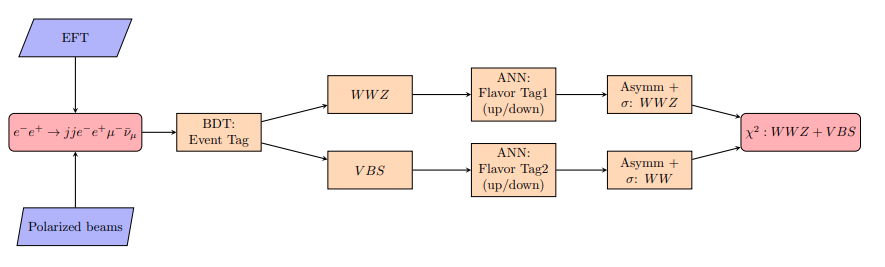}
	  	\caption{Schematic representation of the methodology used in the current analysis. This includes the selection of phase space using boosted decision trees, reconstruction of jet identities with an artificial neural network, and the use of angular distributions for polarization and spin correlation studies. The analysis process also involves Monte Carlo Markov chain techniques to constrain the Wilson coefficients under varying statistical and systematic errors.}
	  	\label{fig:flow}
	  \end{figure*}
      \\
	  It is to further note that some of the polarization (for e.g. $A_y, A_{xy}, \cdots$) and spin correlation asymmetries (e.g. $A_{x,y}^{12}, A_{xy,z}^{12}, A_{x,z,y}^{123},\cdots$) are parity-odd which implies that these spin asymmetries averages out if the final daughter of weak boson are not identified. In the case of leptons as a final products, we assume an ideal identification. Whereas for the two jets resulting from light quarks, the identification becomes non-trivial. To achieve the required classification of jets, we developed a specific tagger using artificial neural network~(ANN). The tagging of two jets are done in two class viz. \emph{up} and \emph{down}-type jets following the decay of $W$ boson. The two different tagger are developed for two tagged events viz., WWZ and VBS-like events. We take two hardest jets with the truth label decided by the geometric distance between the jets~($j$) and the parent light quarks~($q$), $\Delta R_{jq} = \sqrt{\Delta\eta_{jq}^2+\Delta\phi_{jq}^2}$. The features used for tagging are constructed using the constituents of jets (see Appendix`\ref{sec:feature} for features details). The inputs are preprocessed using the {\tt MinMaxScaler} from {\tt sklearn} with default scaling range $[0,1]$. The network architecture consists of three hidden layers with 120, 60, and 30 nodes, respectively, and {\tt ReLU} activation functions. A {\tt Sigmoid} activation is applied to the output node to enable binary classification. The model is trained using binary cross-entropy loss optimized with the {\tt Adam} optimizer. The performance of the network is obtained by testing the randomly selected $40\%$ of test datasets for 1000 iteration. The accuracy on tagging the jets of WWZ events is $\approx 78\%$ and $\approx 70\%$ for the case of VBS events. We show the distribution of accuracy in Fig.~\ref{fig:flavor} for completeness. The trained model is subsequently used for reconstructing spin asymmetries. In the next section, we describe the role of spin related observables on putting a bound on anomalous couplings for different set of luminosities and systematic errors. A flow-chart illustrating the methodology of our analysis is shown in Fig.~\ref{fig:flow} for quick reference.
        \section{Probe of anomalous couplings}
	  \label{sec:probe}
	  The angular distribution provides a large set of observables sensitive to the anomalous couplings in the form of polarizations and spin correlations. The general idea is to obtain as many observables as possible in order to better obtain a distinction between the SM and new physics contribution. As shown in earlier work~\cite{Rahaman:2018ujg}, the asymmetries play a prominent role in parameter extraction given an excess in collider data even if they have minimal contribution on constraining anomalous couplings. Thus large set of polarization and spin correlation asymmetries becomes very important on measurement of the anomalous couplings. To derive a semi-analytical expression for observables in terms of anomalous couplings, we fit the data as follows. Considering the amplitudes present in current analysis (see fig.~\ref{fig:process}) and assuming the order of SMEFT insertion at the amplitude level to be $\le 2$, the cross~section can be expressed as
	  \begin{equation}
	  	\begin{aligned}
	  		\sigma &= \sigma_0 + \sum_{i}\sigma_i c_i + \sum_{i \ge j}\sigma_{ij} c_ic_j + \sum_{i \ge j \ge k}\sigma_{ijk}c_ic_jc_k + \sum_{i \ge j \ge k \ge l}\sigma_{ijkl}c_ic_jc_kc_l.
	  	\end{aligned}
	  \end{equation}
	  Here, the coefficients represent different orders of interaction:
	  \begin{itemize}
            \item $\sigma_0$ is the SM point cross~section.
	  	\item $\sigma_i$ ($\mathcal{O}(\Lambda^{-2})$) corresponds to the interference of SMEFT with the SM.
	  	\item $\sigma_{ij}$ ($\mathcal{O}(\Lambda^{-4})$) are the quadratic terms ($i=j$) for one SMEFT insertion and interference ($i > j$) of two SMEFT insertion.
        \item $\sigma_{ijk}~(\mathcal{O}(\Lambda^{-6}))$ represents the three SMEFT insertion at the cross~section level.
        \item $\sigma_{ijkl}~(\mathcal{O}(\Lambda^{-8}))$ signifies the matrix-squared level of amplitude terms with two insertion of SMEFT.
	  \end{itemize}
	  Dimensionally, $\sigma_{ij}$ correspond to dim-8 operators. From a naive dimension counting perspective, a complete analysis requires these higher-dimensional operators to be accounted. However, due to the suppression of higher-order contributions by larger powers of $\Lambda$, one generally propose a cutoff at $\mathcal{O}(\Lambda^{-4})$. However, cutting off higher dimensions might not be straightforward, as pointed out in Ref.~\cite{Azatov:2016sqh}, where it was noted that due to the non-interference between the SM and $d=6$ operators for amplitudes with at least one transversely polarized boson, the relevant effects might actually come from operators with $d>6$. To substantiate this assumption, we compare different coefficients for a set of $(c_{WWW}, c_W)$ couplings. We obtain that the SM cross-section is of the order of $0.1$~fb, while the coefficients $\sigma_i$ and $\sigma_{ii}$ (for $i=j$) are of the order of $0.001$~fb. The coefficients $\sigma_{ij}$ ($i \neq j$) are found to be of the order of $10^{-6}$~fb, whereas $\sigma_{ijk}$ and $\sigma_{ijkl}$ are suppressed to values below $10^{-9}$~fb. In fact, the magnitudes of $\sigma_{ijk}$ and $\sigma_{ijkl}$ are comparable to or smaller than the intrinsic Monte Carlo statistical uncertainty of the simulations, rendering these contributions effectively indistinguishable from numerical noise. Consequently, they can be safely neglected, and we restrict our analysis to the one-EFT insertion terms, i.e.\ up to $\mathcal{O}(\Lambda^{-4})$.
      \\
       As stated in the previous section, there are respectively $80$ and $728$ asymmetries for VBS and WWZ process which are further binned in the region of transverse momenta of the hardest lepton. We choose seven different bins in a momentum width of $50$~GeV, thus giving us a total of $560$ and $5096$ asymmetries for VBS and WWZ process along with seven binned cross~section, respectively. For each bins, we construct cross~section weighted spin asymmetries and the observables are classified in terms of their $CP$ structure and the corresponding fits are done as,
	  	\begin{align}
	  		&\text{Cross~section}:~ \sigma = \sigma_0 + \sum_{i}\sigma_ic_i + \sum_{i\ge j}\sigma_{ij}c_{i}c_j,\\ 
	  		&\text{CP-Even Asymmetries}:~ A_{\text{Even}} = \frac{\Delta \sigma_E}{\sigma},\\
	  		&\text{CP-Odd Asymmetries}:~ A_{\text{Odd}} = \frac{\Delta\sigma_O }{\sigma},\\
	  		&\Delta\sigma_O= \sigma_ic_i + \sum_{i>j}\sigma_{ij}c_ic_j,\\
	  		&\Delta\sigma_E = \sigma        
	  	\end{align}
       Here, $\Delta\sigma \equiv \sigma A$ denotes the product of cross~section and asymmetries in a particular bin of muon transverse momenta. The binned asymmetries along with the cross~section for two set of beam polarization are used to perform a chi-squared analysis to obtain the bounds on anomalous couplings $c_i$. The chi-squared is defined as,
	  \begin{equation}
	  	\begin{aligned}
	  		\Delta \chi^2(\mathcal{O},c_i,\pm\eta_3,\mp\xi_3) &= \sum_{l,k} \left[\left(\frac{\mathscr{O}_k^l(c_i,+\eta_3,-\xi_3) - \mathscr{O}_k^l(0,+\eta_3,-\xi_3)}{\delta \mathscr{O}^l_k(0,+\eta_3,-\xi_3)}\right)^2 \right.\\ &\left.+  \left(\frac{\mathscr{O}_k^l(c_i,-\eta_3,+\xi_3) - \mathscr{O}_k^l(0,-\eta_3,+\xi_3)}{\delta \mathscr{O}^l_k(0,-\eta_3,+\xi_3)}\right)^2 \right].
	  	\end{aligned}
	  	\label{eqn:chisum}
	  \end{equation}  
	  Here $k,l$ runs over all bins and observables separately, and $\delta\mathcal{O} = \sqrt{(\delta\mathcal{O}_{\text{stat}})^2 + (\delta\mathcal{O}_{\text{syst}})^2}$ is the estimated error in $\mathcal{O}$. For cross~section $\sigma$, the estimated error is given by,
       \begin{equation}
	  	\delta\sigma = \sqrt{\frac{\sigma}{\mathcal{L}}+(\epsilon_\sigma\sigma)^2},
	  	\label{eqn:xsecstat}
	  \end{equation}
	  and if the observable is asymmetry, the error is given by,
	  \begin{equation}
	  	\delta\mathcal{A} = \sqrt{\frac{1-\mathcal{A}^2}{\mathcal{L}\sigma}+\epsilon_\mathcal{A}^2}.
	  \end{equation}
	  Here, $\epsilon_\mathcal{A},\epsilon_\sigma$ are the fractional systematic error in asymmetry and cross~section, respectively, and $\mathcal{L}$ is the integrated luminosity. 
	  
\begin{table*}[!htb]
	\centering
	\caption{\label{tab:Lim0ne}One parameter limits on anomalous couplings~$c_i$ at $95\%$ CL obtained from triple gauge production~$(WWZ)$, vector boson scattering~($VBS$) amplitudes and their combinations. The limits are obtained for $e^-e^+ \to 2je^-e^+l^-\slashed{E}$ process at center of mass energy $\sqrt{s}=1$ TeV with an integrated luminosity $\mathcal{L} = 1000$ fb$^{-1}$ and zero systematic. The two set of initial beam polarization~$(\eta_3 = \mp 0.8, \xi_3 = \pm 0.6)$ are used.}
	\renewcommand{\arraystretch}{1.0}
	\begin{tabular*}{\textwidth}{@{\extracolsep{\fill}}c*{3}{>{}c<{}}@{}}
		\hline \hline
		\multirow{2}{*}{$c_i$}&\multicolumn{3}{c}{Limits~(TeV$^{-2}$)}\\
		\cline{2-4}
		&VBS&WWZ&VBS+WWZ\\   \hline
		$c_B/\Lambda^2$&$[-0.31,+0.37]$&$[-0.63,+0.63]$&$[-0.28,+0.32]$\\
		$c_{W}/\Lambda^2$&$[-0.52,+0.49]$&$[-0.56,+0.71]$&$[-0.39,+0.40]$ \\
		$c_{\widetilde{W}}/\Lambda^2$&$[-2.12,+2.12]$&$[-0.91,+0.91]$&$[-0.90,+0.90]$\\
		$c_{WWW}/\Lambda^2$&$[-0.43,+1.87]$&$[-0.31,+0.43]$&$[-0.26,+0.36]$\\
		$c_{\widetilde{W}WW}/\Lambda^2$&$[-2.27,+2.27]$&$[-0.85,+0.85]$&$[-0.84,+0.84]$\\
		$c_{WW}/\Lambda^2$&$[-0.81,+0.84]$&$[-0.56,+0.55]$&$[-0.48,+0.48]$\\
		$c_{\widetilde{W}W}/\Lambda^2$&$[-1.25,+1.25]$&$[-0.69,+0.69]$&$[-0.66,+0.66]$\\
		$c_{BB}/\Lambda^2$&$[-1.50,+1.57]$&$[-1.51,+1.43]$&$[-1.08,+1.07]$\\
		$c_{\widetilde{B}B}/\Lambda^2$&$[-3.17,+3.17]$&$[-1.54,+1.54]$&$[-1.50,+1.50]$\\
		\hline\hline
	\end{tabular*}
\end{table*}	  
We derive one-parameter bounds on the nine independent dimension-six operators 
contributing to anomalous gauge interactions at 
$\sqrt{s}=1~\text{TeV}$ with initial beam polarization 
$(\eta_3,\xi_3) = (\mp 0.8,\pm 0.6)$. The limits, quoted at the 
$95\%$ confidence level (CL), are obtained by performing a binned $\chi^2$ analysis 
over seven kinematic bins ($p_T$ of hardest lepton) of the $ 2j\,e^-e^+\,\ell^-\slashed{E}$ final state, 
assuming an integrated luminosity of $\mathcal{L}=1000~\text{fb}^{-1}$ and neglecting 
systematic uncertainties. Table~\ref{tab:Lim0ne} summarizes the individual sensitivities 
from vector boson scattering (VBS) and triple gauge boson production (WWZ), as well as 
the improved reach from their statistical combination. 

We observe that the WWZ channel provides the dominant sensitivity for a large subset 
of Wilson coefficients. In particular, the CP-even WCs 
$c_{WWW}$ and $c_{WW}$ are significantly better constrained by 
WWZ compared to VBS. Similarly, the CP-odd WCs 
$c_{\widetilde{W}}$, $c_{\widetilde{W}W}$, 
$c_{\widetilde{W}WW}$ and $c_{\widetilde{B}B}$ all yield markedly 
tighter bounds in the WWZ channel, where the interference terms with the SM amplitude 
are more pronounced. In contrast, the CP-even coefficients 
$c_B$, and $c_W$ are better constrained in the VBS channel. 
The coefficient $c_{BB}$ receives nearly comparable contributions from both 
processes. Importantly, across all operators the statistical combination of VBS and 
WWZ leads to further improvement, reducing the allowed parameter space typically by 
$\mathcal{O}(20\!-\!40\%)$ and in some cases by almost a factor of two. This demonstrates 
the strong complementarity between vector boson scattering and triple gauge boson 
production in probing the complete set of anomalous electroweak interactions at 
future $e^+e^-$ colliders.

\begin{figure*}[!htb]
	  	\centering
	  	\includegraphics[width=0.49\textwidth]{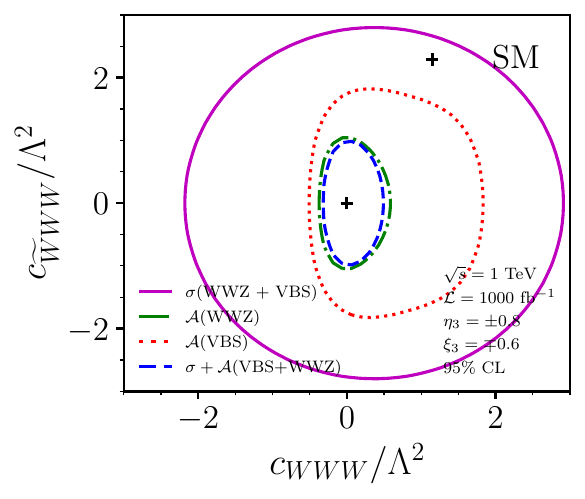}
	  	\includegraphics[width=0.49\textwidth]{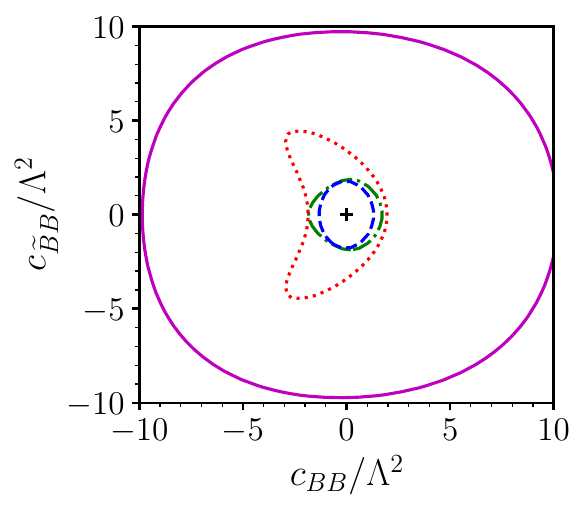}
	  	\includegraphics[width=0.49\textwidth]{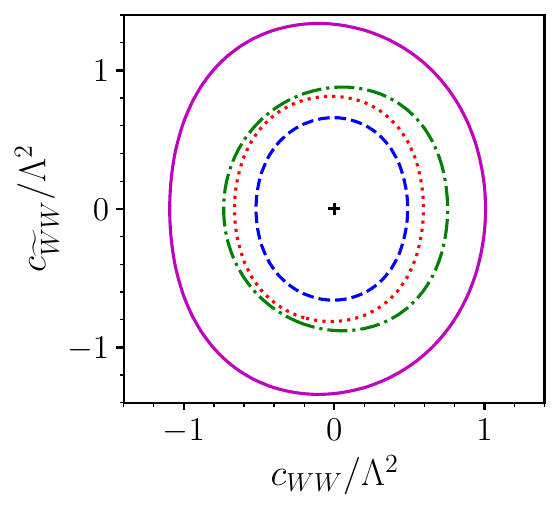}
	  	\includegraphics[width=0.49\textwidth]{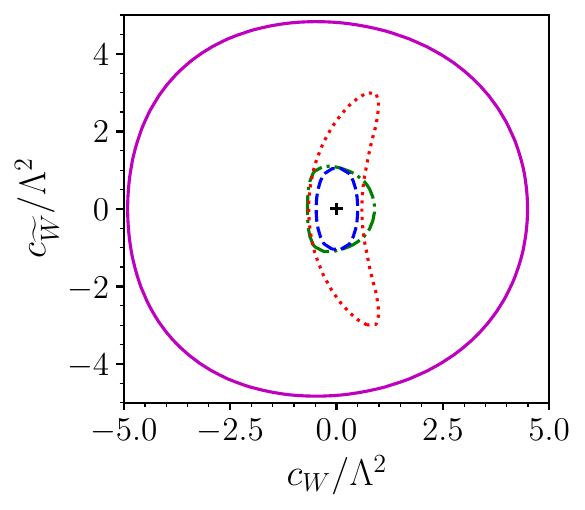}
	  	\caption{\label{fig:twod}Two dimensional $95\%$ CL contours computed with $\Delta\chi^2$ function for cross~section ($\sigma$), polarization and spin correlation asymmetries $(\mathcal{A})$ in all seven bins of $p_T^{\mu^-}$ as a function of two anomalous couplings. The contours are obtain at $\sqrt{s}=1$ TeV, $\mathcal{L}=1000$ fb$^{-1}$, and zero systematics. The legend details remains same to that of left top panel for all figures.}
	  \end{figure*}

Next, we quantify the role of various observables viz. cross~section, polarization and spin correlations of VBS and WWZ amplitudes on setting the limits of anomalous couplings. We achieve this by presenting two dimensional $95\%$ CL contour plots for different observables as a function of two anomalous couplings, while setting others to zero, in Fig.~\ref{fig:twod}. 
We choose pairs $(c_{WWW},c_{\widetilde{W}WW})$ (top left panel), $(c_{WW},c_{\widetilde{W}W})$ (right top panel), $(c_{BB},c_{\widetilde{B}B})$ (left bottom panel), and $(c_W,c_{\widetilde{W}})$ (right bottom panel) to demonstrate the role of various observables. The immediate observation is the least contribution of cross~section (solid magenta curve) to $\Delta\chi^2$ except for $c_{WW}$ coupling (see left top panel) where the least contribution to $c_{WW}$ is provided by asymmetries associated with WWZ process (dashed green curve). Nevertheless, the asymmetries reconstructed from WWZ process gives a dominant contribution for all other couplings followed by that of VBS amplitudes (dotted red curve). The final limits are set by combining cross~section and asymmetries of both processes, which is represented by dashed blue curve. The vertex factor for triple gauge, Higgs-gauge, or quartic couplings in presence of dim-6 operators contains a momentum dependent terms after symmetry breaking. The potential of such terms are fully exploited at hadron collider like LHC which runs at comparatively higher energy than the electron-positron collider. The higher center of mass energy increases the sensitivity of cross~section to various anomalous couplings which leads to tighter constraint using cross~section alone at LHC. Though, the current analysis is performed at the maximum $\sqrt{s}$ (1 TeV) planned at ILC which is around $13$ times lower than that of current LHC (the factor reduces for $\sqrt{\hat{s}}$); the lowering of rate sensitivity is complemented by large number of spin observables.
\\
Finally, we perform multi-dimensional Markov Chain Monte Carlo analysis to obtain the marginalized limits on all nine anomalous couplings for different values of integrated luminosity and systematic errors. The likelihood ratio function for a parameter space, $\vec{x}\equiv (c_i, \eta_3, \xi_3)$, is considered to be $\chi^2$ which is given by Eq.~(\ref{eqn:chisum}),
\begin{equation*}
	-2~\text{ln}(\Delta L(\vec{x})) = \chi^2(\vec{x}).
\end{equation*}
The further interpretation of Monte Carlo samples for Bayesian inference is done using {\tt GetDist} \cite{Lewis:2019xzd}. It is a {\tt Python} based package that deals on estimating from the MC samples which includes confidence limits and marginalized densities. We perform MCMC analysis based on {\tt Metropolis-Hastings}~\cite{Metropolis:1953am,Hastings:1970aa} algorithm by combining all observables from seven bins of muon transverse momenta. The different values of integrated luminosities chosen are,
$$\mathcal{L} = \{100~\text{fb}^{-1},300~\text{fb}^{-1},1000~\text{fb}^{-1},3000~\text{fb}^{-1}\},$$
and systematic errors are,
$$(\epsilon_\sigma,\epsilon_A) = \{(0,0),(1\%,0.5\%),(5\%,5\%),(10\%,10\%)\}.$$
\begin{figure*}[!t]
	\centering
	\includegraphics[width=0.49\textwidth]{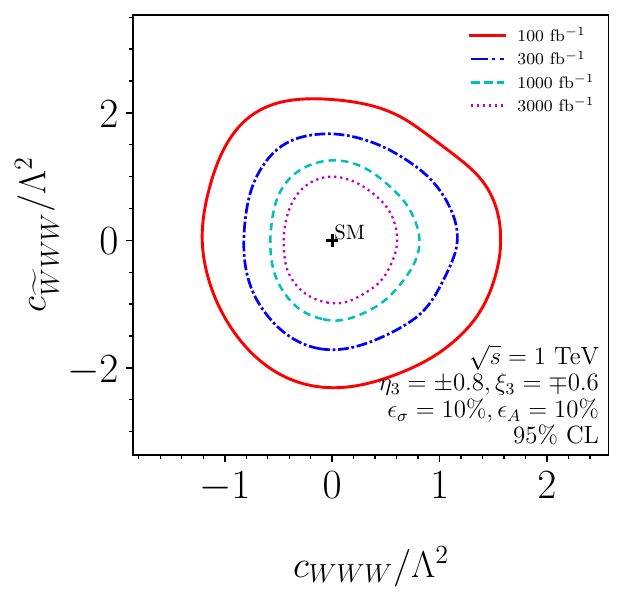}
	\includegraphics[width=0.49\textwidth]{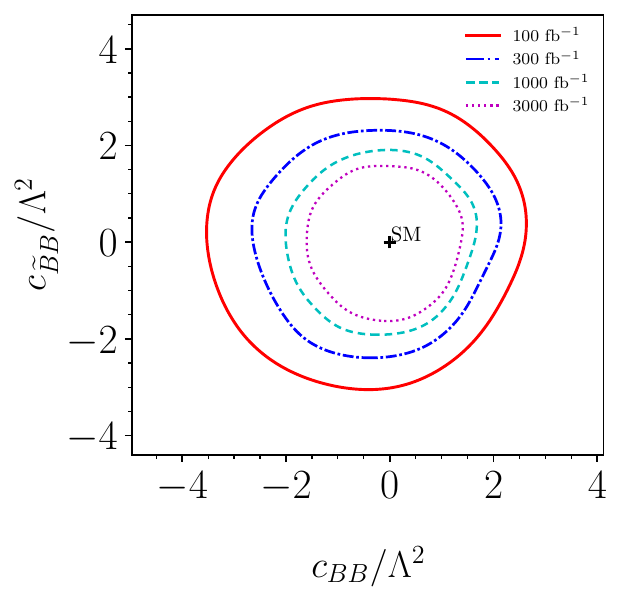}
	\includegraphics[width=0.49\textwidth]{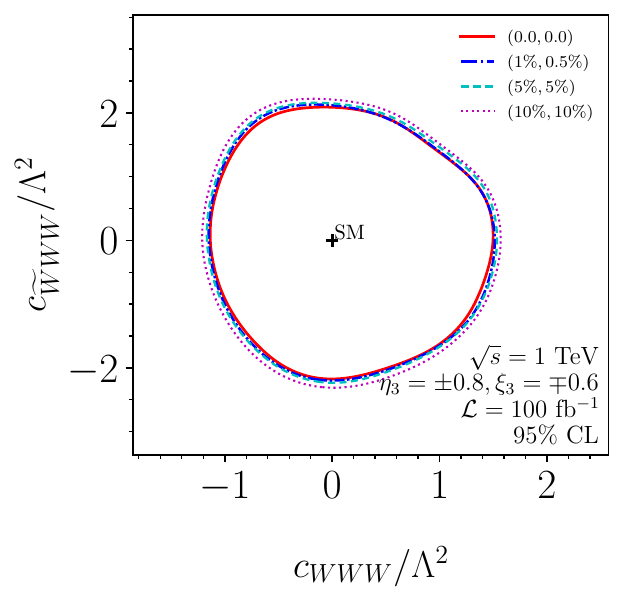}
	\includegraphics[width=0.49\textwidth]{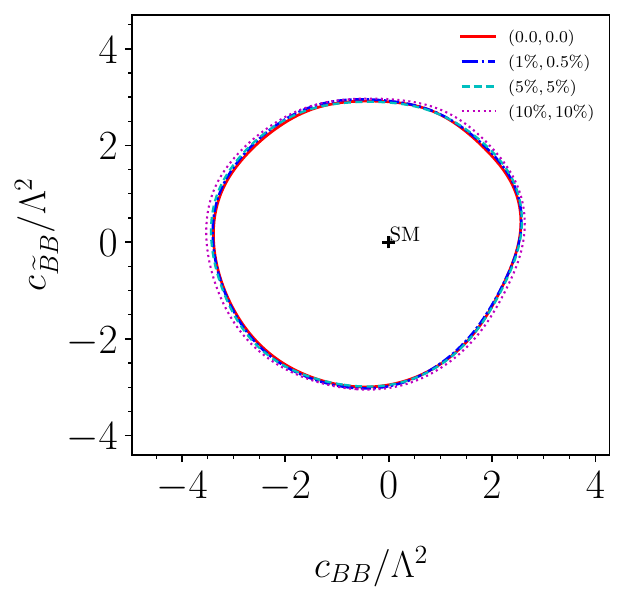}
	\caption{\label{fig:systlumi}$95\%$ CL marginalized 2-D projections on $c_{WWW}-c_{\widetilde{W}WW}$, and $c_{BB}-c_{\widetilde{B}B}$ planes from MCMC for $\sqrt{s}=1$ TeV as a function of integrated luminosity (top row) and systematic errors (bottom row).}
\end{figure*}
We perform a study to quantify the marginalized limits variation of anomalous couplings with respect to luminosity and systematic errors. In Fig.~\ref{fig:systlumi}, we show a 2-D projections at $95\%$ CL for $(c_{WWW}-c_{\widetilde{W}WW})$ and $(c_{BB}-c_{\widetilde{B}B})$ planes as a function of luminosity (top row) and systematic error (bottom row). From the figure, it becomes evident that one improves on precision of probing anomalous effects with increasing datasets. The limits does not show a significant deviation on increasing the systematic error from zero to more conservative value of $10\%$ for both observables. It is due to the low cross~section such that the errors are dominated by statistical ones. Thus, the future electron-positron could be perfect place to perform a precision measurement on multi gauge and Higgs-gauge couplings provided sufficient luminosity is achieved. And finally for completeness, we list the marginalized limits on all nine WCs in Table~\ref{tab:mcmc} for different values of luminosity and systematic errors.
\begin{table*}[!htb]
		\centering
		\caption{\label{tab:mcmc}Posterior 95\% Bayesian confidence intervals for anomalous couplings~$c_i$~(TeV$^{-2}$) obtained from Markov Chain Monte Carlo global fits. Limits are shown for the process $e^-e^+\to  2j\, e^-e^+l^-\slashed{E}$ at $\sqrt{s}=1$~TeV with beam polarization $(\eta_3,\xi_3)=(\pm0.8,\mp0.6)$, for different integrated luminosities and systematic er $c_{\widetilde{W}}$,rors.}
		\renewcommand{\arraystretch}{0.7}
		\begin{tabular*}{\textwidth}{@{\extracolsep{\fill}}*{6}{>{}c<{}}@{}}
			\hline
			$c_i$ (TeV$^{-2})$&$(\epsilon_\sigma,\epsilon_A)$&$100$ fb$^{-1}$&$300$ fb$^{-1}$&$1000$ fb$^{-1}$&$3000$ fb$^{-1}$\\
			\hline
			&$(0.0,0.0)$&$[-1.03,+1.65]$&$[-0.71,+0.92]$&$[-0.43,+0.51]$&$[-0.27,+0.29]$\\
			$c_B/\Lambda^2$&$(1\%,0.5\%)$&$[-1.03,+1.67]$&$[-0.71,+0.92]$&$[-0.43,+0.52]$&$[-0.27,+0.30]$\\
			&$(5\%,5\%)$&$[-1.05,+1.72]$&$[-0.76,+0.98]$&$[-0.48,+0.60]$&$[-0.36,+0.41]$\\
			&$(10\%,10\%)$&$[-1.08,+1.75]$&$[-0.83,+1.12]$&$[-0.61,+0.74]$&$[-0.47,+0.58]$\\
			\hline
			&$(0.0,0.0)$&$[-1.32,+1.85]$&$[-0.92,+1.03]$&$[-0.54,+0.56]$&$[-0.34,+0.32]$\\
			$c_W/\Lambda^2$&$(1\%,0.5\%)$&$[-1.33,+1.92]$&$[-0.90,+1.03]$&$[-0.54,+0.55]$&$[-0.34,+0.32]$\\
			&$(5\%,5\%)$&$[-1.35,+1.92]$&$[-0.96,+1.13]$&$[-0.60,+0.67]$&$[-0.44,+0.45]$\\
			&$(10\%,10\%)$&$[-1.38,+1.94]$&$[-1.03,+1.29]$&$[-0.75,+0.82]$&$[-0.57,+0.63]$\\
			\hline
			&$(0.0,0.0)$&$[-1.73,+1.65]$&$[-1.22,+1.19]$&$[-0.78,+0.78]$&$[-0.53,+0.53]$\\
			$c_{\widetilde{W}}/\Lambda^2$&$(1\%,0.5\%)$&$[-1.73,+1.64]$&$[-1.22,+1.19]$&$[-0.79,+0.79]$&$[-0.53,+0.52]$\\
			&$(5\%,5\%)$&$[-1.75,+1.66]$&$[-1.25,+1.23]$&$[-0.87,+0.88]$&$[-0.66,+0.65]$\\
			&$(10\%,10\%)$&$[-1.78,+1.70]$&$[-1.34,+1.28]$&$[-1.02,+1.01]$&$[-0.83,+0.82]$\\
			\hline
			&$(0.0,0.0)$&$[-0.90,+1.12]$&$[-0.54,+0.77]$&$[-0.32,+0.42]$&$[-0.20,+0.24]$\\
			$c_{WWW}/\Lambda^2$&$(1\%,0.5\%)$&$[-0.91,+1.13]$&$[-0.55,+0.77]$&$[-0.32,+0.43]$&$[-0.20,+0.24]$\\
			&$(5\%,5\%)$&$[-0.92,+1.15]$&$[-0.59,+0.79]$&$[-0.37,+0.49]$&$[-0.26,+0.32]$\\
			&$(10\%,10\%)$&$[-0.96,+1.16]$&$[-0.65,+0.85]$&$[-0.46,+0.59]$&$[-0.36,+0.45]$\\
			\hline
			&$(0.0,0.0)$&$[-1.71,+1.63]$&$[-1.18,+1.12]$&$[-0.74,+0.73]$&$[-0.47,+0.47]$\\
			$c_{\widetilde{W}WW}/\Lambda^2$&$(1\%,0.5\%)$&$[-1.72,+1.63]$&$[-1.18,+1.13]$&$[-0.74,+0.73]$&$[-0.47,+0.48]$\\
			&$(5\%,5\%)$&$[-1.73,+1.66]$&$[-1.22,+1.16]$&$[-0.82,+0.79]$&$[-0.58,+0.58]$\\
			&$(10\%,10\%)$&$[-1.80,+1.70]$&$[-1.32,+1.24]$&$[-0.95,+0.92]$&$[-0.75,+0.73]$\\
			\hline
			&$(0.0,0.0)$&$[-1.11,+1.00]$&$[-0.79,+0.70]$&$[-0.50,+0.45]$&$[-0.31,+0.29]$\\
			$c_{WW}/\Lambda^2$&$(1\%,0.5\%)$&$[-1.13,+1.00]$&$[-0.79,+0.71]$&$[-0.50,+0.46]$&$[-0.31,+0.29]$\\
			&$(5\%,5\%)$&$[-1.12,+1.00]$&$[-0.80,+0.72]$&$[-0.52,+0.48]$&$[-0.35,+0.33]$\\
			&$(10\%,10\%)$&$[-1.14,+1.02]$&$[-0.82,+0.75]$&$[-0.55,+0.53]$&$[-0.38,+0.38]$\\
			\hline
			&$(0.0,0.0)$&$[-1.26,+1.20]$&$[-0.90,+0.87]$&$[-0.60,+0.60]$&$[-0.40,+0.39]$\\
			$c_{\widetilde{W}W}/\Lambda^2$&$(1\%,0.5\%)$&$[-1.28,+1.22]$&$[-0.90,+0.87]$&$[-0.60,+0.60]$&$[-0.40,+0.39]$\\
			&$(5\%,5\%)$&$[-1.29,+1.23]$&$[-0.93,+0.89]$&$[-0.65,+0.65]$&$[-0.47,+0.48]$\\
			&$(10\%,10\%)$&$[-1.32,+1.26]$&$[-0.98,+0.94]$&$[-0.72,+0.71]$&$[-0.57,+0.57]$\\
			\hline
			&$(0.0,0.0)$&$[-2.78,+1.82]$&$[-1.87,+1.43]$&$[-1.16,+1.00]$&$[-0.78,+0.71]$\\
			$c_{BB}/\Lambda^2$&$(1\%,0.5\%)$&$[-2.79,+1.84]$&$[-1.89,+1.43]$&$[-1.16,+1.01]$&$[-0.80,+0.71]$\\
			&$(5\%,5\%)$&$[-2.83,+1.84]$&$[-1.96,+1.47]$&$[-1.30,+1.09]$&$[-0.98,+0.84]$\\
			&$(10\%,10\%)$&$[-2.90,+1.90]$&$[-2.13,+1.55]$&$[-1.59,+1.21]$&$[-1.24,+1.03]$\\
			\hline
			&$(0.0,0.0)$&$[-2.35,+2.26]$&$[-1.72,+1.73]$&$[-1.21,+1.18]$&$[-0.86,+0.80]$\\
			$c_{\widetilde{B}B}/\Lambda^2$&$(1\%,0.5\%)$&$[-2.36,+2.32]$&$[-1.75,+1.75]$&$[-1.23,+1.19]$&$[-0.85,+0.83]$\\
			&$(5\%,5\%)$&$[-2.33,+2.28]$&$[-1.76,+1.75]$&$[-1.32,+1.33]$&$[-1.03,+1.01]$\\
			&$(10\%,10\%)$&$[-2.36,+2.35]$&$[-1.87,+1.83]$&$[-1.49,+1.51]$&$[-1.23,+1.23]$\\
			\hline
			\end{tabular*}
		\end{table*}  
        \section{Conclusion and Discussions}
	\label{sec:conclude}
	The effects of higher order operators are usually translated as anomalous contribution to SM couplings or induce non-SM vertex all together as in neutral triple gauge boson couplings. The single SMEFT operator might also effect different Lorentz structure and in this article we tried to study a set of dimension-6 operators constructed out of Higgs and weak boson states which have an impact on triple-gauge, Higgs-gauge and quartic-gauge couplings. We achieved this by probing a $e^-e^+\to 2j e^-e^+ l^- \bar{\nu}_l$ process at $\sqrt{s}=1$ TeV in presence of polarized beams. We classified the amplitudes in two types as vector boson scattering (VBS) and triple gauge boson (WWZ) production like amplitudes and this classification is obtain by boosted decision trees. We exploit the 2-body and 3-body spin correlations, respectively in VBS and WWZ sub-processes, to constrain the relevant anomalous couplings. The polarization and spin correlation asymmetries of $W^+$ boson are reconstructed by tagging the final jets with artificial neural networks. While for the case of lepton final state, we assume the ideal identification of flavor of those leptons.
	\\
	The importance of asymmetries related to various angular variables of final fermions are highlighted on the 2-D contour plots in Fig.~\ref{fig:twod}, where the majority of contribution to $\chi^2$ comes from asymmetries. The translation to precision are noted in Table~\ref{tab:Lim0ne}. While comparing two sub-processes, we note that for couplings, $c_B$, and $c_W$, VBS provides better precision while for the rest of $7$ Wilson coefficient, WWZ proves to be dominant. It is related to the large number of asymmetries $728$ with WWZ in comparison to $80$ asymmetries associated with two weak boson in VBS. We have not consider the correlation with two forward lepton in VBS in this article.
	\\
	The LHC enjoys larger center of mass energy and larger set of amplitudes due to different interacting particles; thus for a set of operators inducing a momentum dependent anomalous couplings the sensitivity from the differential rate could be increased. Nevertheless, we have shown in the current analysis that the large set of observables obtained from the angular distribution of final fermions could complement the lower sensitivity rate in future electron-positron collider.
	\appendix
	\section*{Appendix}
	\section{Orthonormal $J$ basis}
	\label{sec:ji}
	For completeness, the set of orthonormal $J_i$ matrices constructed out of spin-1 operators are,
	\begin{equation*}
		\begin{aligned}
			S_x &= \sqrt{\frac{1}{2}}\begin{pmatrix}
				0&1&0\\
				1&0&1\\
				0&1&0
			\end{pmatrix},~ S_y = \frac{1}{i \sqrt{2}}\begin{pmatrix}
				0&1&0\\
				-1&0&1\\
				0&-1&0
			\end{pmatrix},
			S_z = \begin{pmatrix}
				1&0&0\\
				0&0&0\\
				0&0&-1
			\end{pmatrix}
		\end{aligned}
	\end{equation*}
	The $J_i^\prime$s are defined as,
	\begin{equation*}
		\begin{aligned}
			J_1 &= \frac{S_x}{2}, ~ J_2 = \frac{S_y}{2}, ~ J_3 = \frac{S_z}{2}~J_4 = \frac{1}{2}(S_x\cdot S_y + S_y \cdot S_x),\\
			J_5 &= \frac{1}{2}(S_x\cdot S_z + S_z \cdot S_x),~J_6 = \frac{1}{2}(S_y\cdot S_z + S_z \cdot S_y),\\
			J_7 &= \frac{1}{2}(S_x\cdot S_x - S_y \cdot S_y),~ J_8 = \sqrt{3}\left(\frac{S_z\cdot S_z}{2} - \frac{\mathbb{I}}{3}\right)
		\end{aligned}
	\end{equation*}
	The $J_i^\prime$s matrices follow Tr$[J_i\cdot J_j] = \delta^j_{i}/2$.
	\section{Normalized decay density matrix}
	\label{sec:Nddm}
	The normalized decay density matrices at the helicity rest frame for spin-1 boson are given by,
	\begin{equation}
		\Gamma(\lambda,\lambda^\prime) = \begin{bmatrix}
			\frac{1+\delta+(1-3\delta)\cos^2\theta+2\alpha\cos\theta}{4}&\frac{\sin\theta(\alpha+(1-3\delta)\cos\theta)}{2\sqrt{2}}e^{i\phi}&(1-3\delta)\frac{1-\cos^2\theta}{4}e^{i2\phi}\\
			\frac{\sin\theta(\alpha+(1-3\delta)\cos\theta)}{2\sqrt{2}}e^{-i\phi}&\delta+(1-3\delta)\frac{\sin^2\theta}{2}&\frac{\sin\theta(\alpha-(1-3\delta)\cos\theta)}{2\sqrt{2}}e^{i\phi}\\
			(1-3\delta)\frac{(1-\cos^2\theta)}{4}e^{-i2\phi}&\frac{\sin\theta(\alpha-(1-3\delta)\cos\theta)}{2\sqrt{2}}e^{-i\phi}&\frac{1+\delta+(1-3\delta)\cos^2\theta}{4}2\alpha\cos\theta
		\end{bmatrix},
	\end{equation}
	respectively with polar $\theta$ and azimuthal angle $\phi$ obtained at the rest frame of the mother particle.   
    \section{Features used for flavor tagging of jets decayed from $W$ boson}
    \label{sec:feature}
    The features we used as input to our neural network in order to classify the two hardest jets decayed from $W$ bosons as either {\tt up} or {\tt down}-type can be divided into two classes, discrete and continuous variables. They are discussed in details in Ref.~\cite{Subba:2023rpm} and we list them here for completeness.
\begin{itemize}
\item Discrete Variables:
\begin{itemize}
    \item Total number~({\tt nlep}), positive leptons~({\tt nl$+$}), negative leptons~({\tt nl$-$});
    \item Total number of visible particles~({\tt nvis});
    \item Total number of charged particles~({\tt nch}), positive charged particles~({\tt nch$+$}), negative charged particles~({\tt nch$-$});
    \item Total number of charged kaons~({\tt nK$+$, nK$-$})$^\star$;
    \item Total number of charged pions~({\tt npi$+$, npi$-$})$^\star$;
    \item Total number of hadrons~({\tt nhad});
    \item Total number of charged hadrons~({\tt nChad}), positively charged hadrons~({\tt nChad$+$}), negative hadrons~({\tt nChad$-$});
    \item Displaced tracks satisfying $p_T > 1.0$ GeV are used. They are binned with respect to the lifetime~($\tau$) in mm of their mother particles:
    \begin{itemize}
        \item {\tt c1}: $\tau < 3.0$ and $\tau > 0.3$,
        \item {\tt c2}: $\tau < 30.0$ and $\tau > 3.0$,
        \item {\tt c3}: $\tau < 300.0$ and $\tau > 30.0$,
        \item {\tt c4}: $\tau < 1200.0$ and $\tau > 300.0$,
        \item {\tt c5}: $\tau > 1200.0$.
    \end{itemize}
\item Total number of $+$ve~({\tt pcl}) and $-$ve~({\tt ncl}) mother particles are also counted. The particle that decay and produces secondary displaced vertex are considered. 
    \end{itemize}
\item Continuous Variables:
\begin{itemize}
    \item Energy of photons~({\tt egamma});
    \item $P^{\mu}_i = \sum_{j \in i}p^\mu_j$, $ P_i = \sum_{j \in i}|\vec{p}_j|$ , $|(P^\mu_i)_T|$,\\ $i \in \{\text{Leptons}, K^+, K^-, \pi^+, \pi^-,K^0_L,\text{Hadrons}\}$;
    \item Energy of charged hadrons~($E^+_{\text{Had}},E^-_{\text{Had}}$). 
\end{itemize}
\end{itemize}
\acknowledgments

 A. Subba acknowledges the support of ANRF grant CRG/2023/000580.
\bibliographystyle{JHEP} 
\bibliography{refer.bib}

\providecommand{\href}[2]{#2}\begingroup\raggedright\begin{thebibliography}{100}

\bibitem{L3:2004lwm}
{\scshape L3} collaboration, \emph{{Measurement of the cross section of W-boson
  pair production at LEP}},
  \href{https://doi.org/10.1016/j.physletb.2004.08.060}{\emph{Phys. Lett. B}
  {\bfseries 600} (2004) 22}
  [\href{https://arxiv.org/abs/hep-ex/0409016}{{\ttfamily hep-ex/0409016}}].

\bibitem{L3:1999znj}
{\scshape L3} collaboration, \emph{{Measurement of mass and width of the $W$
  boson at LEP}},
  \href{https://doi.org/10.1016/S0370-2693(99)00348-2}{\emph{Phys. Lett. B}
  {\bfseries 454} (1999) 386}
  [\href{https://arxiv.org/abs/hep-ex/9909010}{{\ttfamily hep-ex/9909010}}].

\bibitem{L3:1999cbg}
{\scshape L3} collaboration, \emph{{Direct observation of longitudinally
  polarized $W^\pm$ bosons}},
  \href{https://doi.org/10.1016/S0370-2693(99)01489-6}{\emph{Phys. Lett. B}
  {\bfseries 474} (2000) 194}
  [\href{https://arxiv.org/abs/hep-ex/0001016}{{\ttfamily hep-ex/0001016}}].

\bibitem{L3:1998qun}
{\scshape L3} collaboration, \emph{{Measurement of $W$ pair cross-sections in
  $e^{+} e^{-}$ interactions at $S^{(1/2)}$ = 183-GeV and $W$ decay branching
  fractions}}, \href{https://doi.org/10.1016/S0370-2693(98)01036-3}{\emph{Phys.
  Lett. B} {\bfseries 436} (1998) 437}.

\bibitem{L3:2004ulv}
{\scshape L3} collaboration, \emph{{Measurement of triple gauge boson couplings
  of the $W$ boson at LEP}},
  \href{https://doi.org/10.1016/j.physletb.2004.02.045}{\emph{Phys. Lett. B}
  {\bfseries 586} (2004) 151}
  [\href{https://arxiv.org/abs/hep-ex/0402036}{{\ttfamily hep-ex/0402036}}].

\bibitem{ALEPH:1999jcv}
{\scshape ALEPH} collaboration, \emph{{Measurement of W pair production in e+
  e- collisions at 183-GeV}},
  \href{https://doi.org/10.1016/S0370-2693(99)00304-4}{\emph{Phys. Lett. B}
  {\bfseries 453} (1999) 107}
  [\href{https://arxiv.org/abs/hep-ex/9903053}{{\ttfamily hep-ex/9903053}}].

\bibitem{ALEPH:1999ljd}
{\scshape ALEPH} collaboration, \emph{{Measurement of the $W$ mass in $e^{+}
  e^{-}$ collisions at 183-GeV}},
  \href{https://doi.org/10.1016/S0370-2693(99)00305-6}{\emph{Phys. Lett. B}
  {\bfseries 453} (1999) 121}.

\bibitem{ALEPH:2004klc}
{\scshape ALEPH} collaboration, \emph{{Improved measurement of the triple
  gauge-boson couplings gamma W W and Z W W in e+ e- collisions}},
  \href{https://doi.org/10.1016/j.physletb.2005.03.058}{\emph{Phys. Lett. B}
  {\bfseries 614} (2005) 7}.

\bibitem{DELPHI:1999pkm}
{\scshape DELPHI} collaboration, \emph{{Measurements of the trilinear gauge
  boson couplings W W V (V = gamma,Z) in e+ e- collisions at 183-GeV}},
  \href{https://doi.org/10.1016/S0370-2693(99)00671-1}{\emph{Phys. Lett. B}
  {\bfseries 459} (1999) 382}.

\bibitem{DELPHI:1999ypr}
{\scshape DELPHI} collaboration, \emph{{W pair production cross-section and W
  branching fractions in e+ e- interactions at 183-GeV}},
  \href{https://doi.org/10.1016/S0370-2693(99)00488-8}{\emph{Phys. Lett. B}
  {\bfseries 456} (1999) 310}.

\bibitem{DELPHI:1997vbx}
{\scshape DELPHI} collaboration, \emph{{Measurement and interpretation of the
  $W$ pair cross-section in $e^{+} e^{-}$ interactions at 161-GeV}},
  \href{https://doi.org/10.1016/S0370-2693(97)00226-8}{\emph{Phys. Lett. B}
  {\bfseries 397} (1997) 158}.

\bibitem{DELPHI:1997ytm}
{\scshape DELPHI} collaboration, \emph{{Measurement of the W pair cross-section
  and of the W mass in e+ e- interactions at 172-GeV}},
  \href{https://doi.org/10.1007/s100520050163}{\emph{Eur. Phys. J. C}
  {\bfseries 2} (1998) 581}.

\bibitem{OPAL:1997yzg}
{\scshape OPAL} collaboration, \emph{{Measurement of triple gauge boson
  couplings from $W^{+} W^{-}$ production at $S^{(1/2)}$ = 172-GeV}},
  \href{https://doi.org/10.1007/s100520050164}{\emph{Eur. Phys. J. C}
  {\bfseries 2} (1998) 597}
  [\href{https://arxiv.org/abs/hep-ex/9709023}{{\ttfamily hep-ex/9709023}}].

\bibitem{OPAL:1998ixj}
{\scshape OPAL} collaboration, \emph{{$W^{+} W^{-}$ production and triple gauge
  boson couplings at LEP energies up to 183-GeV}},
  \href{https://doi.org/10.1007/s100529901106}{\emph{Eur. Phys. J. C}
  {\bfseries 8} (1999) 191}
  [\href{https://arxiv.org/abs/hep-ex/9811028}{{\ttfamily hep-ex/9811028}}].

\bibitem{OPAL:1998riq}
{\scshape OPAL} collaboration, \emph{{Measurement of the $W$ mass and width in
  $e^{+} e^{-}$ collisions at 183-GeV}},
  \href{https://doi.org/10.1016/S0370-2693(99)00308-1}{\emph{Phys. Lett. B}
  {\bfseries 453} (1999) 138}
  [\href{https://arxiv.org/abs/hep-ex/9901025}{{\ttfamily hep-ex/9901025}}].

\bibitem{OPAL:2003xqq}
{\scshape OPAL} collaboration, \emph{{Measurement of charged current triple
  gauge boson couplings using $W$ pairs at LEP}},
  \href{https://doi.org/10.1140/epjc/s2003-01524-6}{\emph{Eur. Phys. J. C}
  {\bfseries 33} (2004) 463}
  [\href{https://arxiv.org/abs/hep-ex/0308067}{{\ttfamily hep-ex/0308067}}].

\bibitem{D0:2004fqq}
{\scshape D0} collaboration, \emph{{Measurement of the $W W$ production cross
  section in $p\bar{p}$ collisions at $\sqrt{s} = 1.96$ TeV}},
  \href{https://doi.org/10.1103/PhysRevLett.100.139901}{\emph{Phys. Rev. Lett.}
  {\bfseries 94} (2005) 151801}
  [\href{https://arxiv.org/abs/hep-ex/0410066}{{\ttfamily hep-ex/0410066}}].

\bibitem{D0:2006eed}
{\scshape D0} collaboration, \emph{{Limits on anomalous trilinear gauge
  couplings from $W W \to e^{+} e^{-}$, $W W \to e^\pm \mu^\mp$, and $W W \to
  \mu^{+} \mu^{-}$ events from $p\bar{p}$ collisions at $\sqrt{s}$ =
  1.96-TeV}}, \href{https://doi.org/10.1103/PhysRevD.74.057101}{\emph{Phys.
  Rev. D} {\bfseries 74} (2006) 057101}
  [\href{https://arxiv.org/abs/hep-ex/0608011}{{\ttfamily hep-ex/0608011}}].

\bibitem{D0:2009xgd}
{\scshape D0} collaboration, \emph{{Measurement of the WW production cross
  section with dilepton final states in p anti-p collisions at s**(1/2) =
  1.96-TeV and limits on anomalous trilinear gauge couplings}},
  \href{https://doi.org/10.1103/PhysRevLett.103.191801}{\emph{Phys. Rev. Lett.}
  {\bfseries 103} (2009) 191801}
  [\href{https://arxiv.org/abs/0904.0673}{{\ttfamily 0904.0673}}].

\bibitem{CDF:1995vgr}
{\scshape CDF} collaboration, \emph{{Limits on $W W Z$ and $W W \gamma$
  couplings from $W W$ and $W Z$ production in $p \bar{p}$ collisions at
  $\sqrt{s} = 1.8$ TeV}},
  \href{https://doi.org/10.1103/PhysRevLett.75.1017}{\emph{Phys. Rev. Lett.}
  {\bfseries 75} (1995) 1017}
  [\href{https://arxiv.org/abs/hep-ex/9503009}{{\ttfamily hep-ex/9503009}}].

\bibitem{CDF:2005xet}
{\scshape CDF} collaboration, \emph{{Measurement of the $W^+ W^-$ production
  cross section in $p\bar{p}$ collisions at $\sqrt{s} = 1.96$ TeV using
  dilepton events}},
  \href{https://doi.org/10.1103/PhysRevLett.94.211801}{\emph{Phys. Rev. Lett.}
  {\bfseries 94} (2005) 211801}
  [\href{https://arxiv.org/abs/hep-ex/0501050}{{\ttfamily hep-ex/0501050}}].

\bibitem{CDF:2007aqs}
{\scshape CDF} collaboration, \emph{{Limits on Anomalous Triple Gauge Couplings
  in $p \bar{p}$ Collisions at $\sqrt{s}$ = 1.96-TeV}},
  \href{https://doi.org/10.1103/PhysRevD.76.111103}{\emph{Phys. Rev. D}
  {\bfseries 76} (2007) 111103}
  [\href{https://arxiv.org/abs/0705.2247}{{\ttfamily 0705.2247}}].

\bibitem{ATLAS:2011vfj}
{\scshape ATLAS} collaboration, \emph{{Measurement of the $W W$ cross section
  in $\sqrt{s}=7$ TeV $pp$ collisions with ATLAS}},
  \href{https://doi.org/10.1103/PhysRevLett.107.041802}{\emph{Phys. Rev. Lett.}
  {\bfseries 107} (2011) 041802}
  [\href{https://arxiv.org/abs/1104.5225}{{\ttfamily 1104.5225}}].

\bibitem{ATLAS:2012mec}
{\scshape ATLAS} collaboration, \emph{{Measurement of $W^+W^-$ production in pp
  collisions at $\sqrt{s}$=7 TeV with the ATLAS detector and limits on
  anomalous WWZ and WW\ensuremath{\gamma} couplings}},
  \href{https://doi.org/10.1103/PhysRevD.87.112001}{\emph{Phys. Rev. D}
  {\bfseries 87} (2013) 112001}
  [\href{https://arxiv.org/abs/1210.2979}{{\ttfamily 1210.2979}}].

\bibitem{ATLAS:2012upi}
{\scshape ATLAS} collaboration, \emph{{Measurement of the $W W$ cross section
  in $\sqrt{s}=7$ TeV $pp$ collisions with the ATLAS detector and limits on
  anomalous gauge couplings}},
  \href{https://doi.org/10.1016/j.physletb.2012.05.003}{\emph{Phys. Lett. B}
  {\bfseries 712} (2012) 289}
  [\href{https://arxiv.org/abs/1203.6232}{{\ttfamily 1203.6232}}].

\bibitem{ATLAS:2016bkj}
{\scshape ATLAS} collaboration, \emph{{Measurements of $W^\pm Z$ production
  cross sections in $pp$ collisions at $\sqrt{s} = 8$ TeV with the ATLAS
  detector and limits on anomalous gauge boson self-couplings}},
  \href{https://doi.org/10.1103/PhysRevD.93.092004}{\emph{Phys. Rev. D}
  {\bfseries 93} (2016) 092004}
  [\href{https://arxiv.org/abs/1603.02151}{{\ttfamily 1603.02151}}].

\bibitem{ATLAS:2016nmw}
{\scshape ATLAS} collaboration, \emph{{Search for anomalous electroweak
  production of $WW/WZ$ in association with a high-mass dijet system in $pp$
  collisions at $\sqrt{s}=8$ TeV with the ATLAS detector}},
  \href{https://doi.org/10.1103/PhysRevD.95.032001}{\emph{Phys. Rev. D}
  {\bfseries 95} (2017) 032001}
  [\href{https://arxiv.org/abs/1609.05122}{{\ttfamily 1609.05122}}].

\bibitem{ATLAS:2016zwm}
{\scshape ATLAS} collaboration, \emph{{Measurement of total and differential
  $W^+W^-$ production cross sections in proton-proton collisions at $\sqrt{s}=$
  8 TeV with the ATLAS detector and limits on anomalous triple-gauge-boson
  couplings}}, \href{https://doi.org/10.1007/JHEP09(2016)029}{\emph{JHEP}
  {\bfseries 09} (2016) 029}
  [\href{https://arxiv.org/abs/1603.01702}{{\ttfamily 1603.01702}}].

\bibitem{ATLAS:2017bbg}
{\scshape ATLAS} collaboration, \emph{{Measurement of the $W^+W^-$ production
  cross section in $pp$ collisions at a centre-of-mass energy of $\sqrt{s}$ =
  13 TeV with the ATLAS experiment}},
  \href{https://doi.org/10.1016/j.physletb.2017.08.047}{\emph{Phys. Lett. B}
  {\bfseries 773} (2017) 354}
  [\href{https://arxiv.org/abs/1702.04519}{{\ttfamily 1702.04519}}].

\bibitem{ATLAS:2018mxa}
{\scshape ATLAS} collaboration, \emph{{Observation of electroweak $W^{\pm}Z$
  boson pair production in association with two jets in $pp$ collisions at
  $\sqrt{s} =$ 13 TeV with the ATLAS detector}},
  \href{https://doi.org/10.1016/j.physletb.2019.05.012}{\emph{Phys. Lett. B}
  {\bfseries 793} (2019) 469}
  [\href{https://arxiv.org/abs/1812.09740}{{\ttfamily 1812.09740}}].

\bibitem{ATLAS:2019rob}
{\scshape ATLAS} collaboration, \emph{{Measurement of fiducial and differential
  $W^+W^-$ production cross-sections at $\sqrt{s}=13$ TeV with the ATLAS
  detector}}, \href{https://doi.org/10.1140/epjc/s10052-019-7371-6}{\emph{Eur.
  Phys. J. C} {\bfseries 79} (2019) 884}
  [\href{https://arxiv.org/abs/1905.04242}{{\ttfamily 1905.04242}}].

\bibitem{CMS:2013ant}
{\scshape CMS} collaboration, \emph{{Measurement of the $W^+W^-$ Cross Section
  in $pp$ Collisions at $\sqrt{s} = 7$ TeV and Limits on Anomalous $WW\gamma$
  and $WWZ$ Couplings}},
  \href{https://doi.org/10.1140/epjc/s10052-013-2610-8}{\emph{Eur. Phys. J. C}
  {\bfseries 73} (2013) 2610}
  [\href{https://arxiv.org/abs/1306.1126}{{\ttfamily 1306.1126}}].

\bibitem{CMS:2011egr}
{\scshape CMS} collaboration, \emph{{Measurement of $W^+ W^-$ production and
  search for the Higgs boson in pp collisions at $\sqrt s=7$ TeV}},
  \href{https://doi.org/10.1016/j.physletb.2011.03.056}{\emph{Phys. Lett. B}
  {\bfseries 699} (2011) 25} [\href{https://arxiv.org/abs/1102.5429}{{\ttfamily
  1102.5429}}].

\bibitem{CMS:2015tmu}
{\scshape CMS} collaboration, \emph{{Measurement of the ${{\mathrm{W} }^{+}
  }\mathrm{W}^{-} $ cross section in pp collisions at $\sqrt{s} =$ 8 TeV and
  limits on anomalous gauge couplings}},
  \href{https://doi.org/10.1140/epjc/s10052-016-4219-1}{\emph{Eur. Phys. J. C}
  {\bfseries 76} (2016) 401}
  [\href{https://arxiv.org/abs/1507.03268}{{\ttfamily 1507.03268}}].

\bibitem{CMS:2013piy}
{\scshape CMS} collaboration, \emph{{Measurement of $W^+ W^-$ and $ZZ$
  Production Cross Sections in $pp$ Collisions at $\sqrt{s} = 8 TeV$}},
  \href{https://doi.org/10.1016/j.physletb.2013.03.027}{\emph{Phys. Lett. B}
  {\bfseries 721} (2013) 190}
  [\href{https://arxiv.org/abs/1301.4698}{{\ttfamily 1301.4698}}].

\bibitem{CMS:2020mxy}
{\scshape CMS} collaboration, \emph{{W$^+$W$^-$ boson pair production in
  proton-proton collisions at $\sqrt{s} =$ 13 TeV}},
  \href{https://doi.org/10.1103/PhysRevD.102.092001}{\emph{Phys. Rev. D}
  {\bfseries 102} (2020) 092001}
  [\href{https://arxiv.org/abs/2009.00119}{{\ttfamily 2009.00119}}].

\bibitem{ATLAS:2014jzl}
{\scshape ATLAS} collaboration, \emph{{Evidence for Electroweak Production of
  $W^{\pm}W^{\pm}jj$ in $pp$ Collisions at $\sqrt{s}=8$ TeV with the ATLAS
  Detector}}, \href{https://doi.org/10.1103/PhysRevLett.113.141803}{\emph{Phys.
  Rev. Lett.} {\bfseries 113} (2014) 141803}
  [\href{https://arxiv.org/abs/1405.6241}{{\ttfamily 1405.6241}}].

\bibitem{ATLAS:2019cbr}
{\scshape ATLAS} collaboration, \emph{{Observation of electroweak production of
  a same-sign $W$ boson pair in association with two jets in $pp$ collisions at
  $\sqrt{s}=13$ TeV with the ATLAS detector}},
  \href{https://doi.org/10.1103/PhysRevLett.123.161801}{\emph{Phys. Rev. Lett.}
  {\bfseries 123} (2019) 161801}
  [\href{https://arxiv.org/abs/1906.03203}{{\ttfamily 1906.03203}}].

\bibitem{ATLAS:2019thr}
{\scshape ATLAS} collaboration, \emph{{Search for the electroweak diboson
  production in association with a high-mass dijet system in semileptonic final
  states in $pp$ collisions at $\sqrt{s}=13$ TeV with the ATLAS detector}},
  \href{https://doi.org/10.1103/PhysRevD.100.032007}{\emph{Phys. Rev. D}
  {\bfseries 100} (2019) 032007}
  [\href{https://arxiv.org/abs/1905.07714}{{\ttfamily 1905.07714}}].

\bibitem{CMS:2020gfh}
{\scshape CMS} collaboration, \emph{{Measurements of production cross sections
  of WZ and same-sign WW boson pairs in association with two jets in
  proton-proton collisions at $\sqrt{s} =$ 13 TeV}},
  \href{https://doi.org/10.1016/j.physletb.2020.135710}{\emph{Phys. Lett. B}
  {\bfseries 809} (2020) 135710}
  [\href{https://arxiv.org/abs/2005.01173}{{\ttfamily 2005.01173}}].

\bibitem{CMS:2014mra}
{\scshape CMS} collaboration, \emph{{Study of vector boson scattering and
  search for new physics in events with two same-sign leptons and two jets}},
  \href{https://doi.org/10.1103/PhysRevLett.114.051801}{\emph{Phys. Rev. Lett.}
  {\bfseries 114} (2015) 051801}
  [\href{https://arxiv.org/abs/1410.6315}{{\ttfamily 1410.6315}}].

\bibitem{CMS:2017fhs}
{\scshape CMS} collaboration, \emph{{Observation of electroweak production of
  same-sign W boson pairs in the two jet and two same-sign lepton final state
  in proton-proton collisions at $\sqrt{s} = $ 13 TeV}},
  \href{https://doi.org/10.1103/PhysRevLett.120.081801}{\emph{Phys. Rev. Lett.}
  {\bfseries 120} (2018) 081801}
  [\href{https://arxiv.org/abs/1709.05822}{{\ttfamily 1709.05822}}].

\bibitem{CMS:2019uys}
{\scshape CMS} collaboration, \emph{{Measurement of electroweak WZ boson
  production and search for new physics in WZ + two jets events in pp
  collisions at $\sqrt{s} =$ 13TeV}},
  \href{https://doi.org/10.1016/j.physletb.2019.05.042}{\emph{Phys. Lett. B}
  {\bfseries 795} (2019) 281}
  [\href{https://arxiv.org/abs/1901.04060}{{\ttfamily 1901.04060}}].

\bibitem{CMS:2019qfk}
{\scshape CMS} collaboration, \emph{{Search for anomalous electroweak
  production of vector boson pairs in association with two jets in
  proton-proton collisions at 13 TeV}},
  \href{https://doi.org/10.1016/j.physletb.2019.134985}{\emph{Phys. Lett. B}
  {\bfseries 798} (2019) 134985}
  [\href{https://arxiv.org/abs/1905.07445}{{\ttfamily 1905.07445}}].

\bibitem{CMS:2020hjs}
{\scshape CMS} collaboration, \emph{{Observation of the Production of Three
  Massive Gauge Bosons at $\sqrt {s}$ =13 TeV}},
  \href{https://doi.org/10.1103/PhysRevLett.125.151802}{\emph{Phys. Rev. Lett.}
  {\bfseries 125} (2020) 151802}
  [\href{https://arxiv.org/abs/2006.11191}{{\ttfamily 2006.11191}}].

\bibitem{ATLAS:2019dny}
{\scshape ATLAS} collaboration, \emph{{Evidence for the production of three
  massive vector bosons with the ATLAS detector}},
  \href{https://doi.org/10.1016/j.physletb.2019.134913}{\emph{Phys. Lett. B}
  {\bfseries 798} (2019) 134913}
  [\href{https://arxiv.org/abs/1903.10415}{{\ttfamily 1903.10415}}].

\bibitem{ATLAS:2016jeu}
{\scshape ATLAS} collaboration, \emph{{Search for triboson $W^{\pm }W^{\pm
  }W^{\mp }$ production in $pp$ collisions at $\sqrt{s}=8$ $\text {TeV}$ with
  the ATLAS detector}},
  \href{https://doi.org/10.1140/epjc/s10052-017-4692-1}{\emph{Eur. Phys. J. C}
  {\bfseries 77} (2017) 141}
  [\href{https://arxiv.org/abs/1610.05088}{{\ttfamily 1610.05088}}].

\bibitem{CMS:2019mpq}
{\scshape CMS} collaboration, \emph{{Search for the production of
  W$^\pm$W$^\pm$W$^\mp$ events at $\sqrt{s} =$ 13 TeV}},
  \href{https://doi.org/10.1103/PhysRevD.100.012004}{\emph{Phys. Rev. D}
  {\bfseries 100} (2019) 012004}
  [\href{https://arxiv.org/abs/1905.04246}{{\ttfamily 1905.04246}}].

\bibitem{Englert:1964et}
F.~Englert and R.~Brout, \emph{{Broken Symmetry and the Mass of Gauge Vector
  Mesons}}, \href{https://doi.org/10.1103/PhysRevLett.13.321}{\emph{Phys. Rev.
  Lett.} {\bfseries 13} (1964) 321}.

\bibitem{Higgs:1964pj}
P.~W. Higgs, \emph{{Broken Symmetries and the Masses of Gauge Bosons}},
  \href{https://doi.org/10.1103/PhysRevLett.13.508}{\emph{Phys. Rev. Lett.}
  {\bfseries 13} (1964) 508}.

\bibitem{Guralnik:1964eu}
G.~S. Guralnik, C.~R. Hagen and T.~W.~B. Kibble, \emph{{Global Conservation
  Laws and Massless Particles}},
  \href{https://doi.org/10.1103/PhysRevLett.13.585}{\emph{Phys. Rev. Lett.}
  {\bfseries 13} (1964) 585}.

\bibitem{ATLAS:2012yve}
{\scshape ATLAS} collaboration, \emph{{Observation of a new particle in the
  search for the Standard Model Higgs boson with the ATLAS detector at the
  LHC}}, \href{https://doi.org/10.1016/j.physletb.2012.08.020}{\emph{Phys.
  Lett. B} {\bfseries 716} (2012) 1}
  [\href{https://arxiv.org/abs/1207.7214}{{\ttfamily 1207.7214}}].

\bibitem{CMS:2012qbp}
{\scshape CMS} collaboration, \emph{{Observation of a New Boson at a Mass of
  125 GeV with the CMS Experiment at the LHC}},
  \href{https://doi.org/10.1016/j.physletb.2012.08.021}{\emph{Phys. Lett. B}
  {\bfseries 716} (2012) 30} [\href{https://arxiv.org/abs/1207.7235}{{\ttfamily
  1207.7235}}].

\bibitem{ATLAS:2013mma}
{\scshape ATLAS} collaboration, \emph{{Combined measurements of the mass and
  signal strength of the Higgs-like boson with the ATLAS detector using up to
  25 fb$^{-1}$ of proton-proton collision data}}, .

\bibitem{Diglio:2014vpa}
S.~Diglio, \emph{{Search for a high mass Higgs boson using the ATLAS
  detector}}, \href{https://doi.org/10.1051/epjconf/20147100038}{\emph{EPJ Web
  Conf.} {\bfseries 71} (2014) 00038}.

\bibitem{CMS:2012qwq}
{\scshape CMS} collaboration, \emph{{Combination of standard model Higgs boson
  searches and measurements of the properties of the new boson with a mass near
  125 GeV}}, .

\bibitem{CMS:2013uhr}
{\scshape CMS} collaboration, \emph{{Updated measurements of the Higgs boson at
  125 GeV in the two photon decay channel}}, .

\bibitem{ATLAS:2024vxc}
{\scshape ATLAS} collaboration, \emph{{Determination of the relative sign of
  the Higgs boson couplings to $W$ and $Z$ bosons using $WH$ production via
  vector-boson fusion with the ATLAS detector}},
  \href{https://arxiv.org/abs/2402.00426}{{\ttfamily 2402.00426}}.

\bibitem{ATLAS:2023owm}
{\scshape ATLAS} collaboration, \emph{{Measurement of the Higgs boson mass with
  $H\to \gamma\gamma$ decays in 140 fb$^{-1}$ of s=13 TeV pp collisions with
  the ATLAS detector}},
  \href{https://doi.org/10.1016/j.physletb.2023.138315}{\emph{Phys. Lett. B}
  {\bfseries 847} (2023) 138315}
  [\href{https://arxiv.org/abs/2308.07216}{{\ttfamily 2308.07216}}].

\bibitem{ATLAS:2023yqk}
{\scshape ATLAS, CMS} collaboration, \emph{{Evidence for the Higgs Boson Decay
  to a Z Boson and a Photon at the LHC}},
  \href{https://doi.org/10.1103/PhysRevLett.132.021803}{\emph{Phys. Rev. Lett.}
  {\bfseries 132} (2024) 021803}
  [\href{https://arxiv.org/abs/2309.03501}{{\ttfamily 2309.03501}}].

\bibitem{CMS:2023gjz}
{\scshape CMS} collaboration, \emph{{Measurements of inclusive and differential
  cross sections for the Higgs boson production and decay to four-leptons in
  proton-proton collisions at $ \sqrt{s} $ = 13 TeV}},
  \href{https://doi.org/10.1007/JHEP08(2023)040}{\emph{JHEP} {\bfseries 08}
  (2023) 040} [\href{https://arxiv.org/abs/2305.07532}{{\ttfamily
  2305.07532}}].

\bibitem{CMS:2023sdw}
{\scshape CMS} collaboration, \emph{{A search for decays of the Higgs boson to
  invisible particles in events with a top-antitop quark pair or a vector boson
  in proton-proton collisions at $\sqrt{s} = 13\,\text {Te}\hspace{-.08em}\text
  {V} $}}, \href{https://doi.org/10.1140/epjc/s10052-023-11952-7}{\emph{Eur.
  Phys. J. C} {\bfseries 83} (2023) 933}
  [\href{https://arxiv.org/abs/2303.01214}{{\ttfamily 2303.01214}}].

\bibitem{ATLAS:2022tnm}
{\scshape ATLAS} collaboration, \emph{{Measurement of the properties of Higgs
  boson production at $\sqrt{s} = 13$ TeV in the $H\to\gamma\gamma$ channel
  using $139$ fb$^{-1}$ of $pp$ collision data with the ATLAS experiment}},
  \href{https://doi.org/10.1007/JHEP07(2023)088}{\emph{JHEP} {\bfseries 07}
  (2023) 088} [\href{https://arxiv.org/abs/2207.00348}{{\ttfamily
  2207.00348}}].

\bibitem{CMS:2022wpo}
{\scshape CMS} collaboration, \emph{{Measurement of the Higgs boson inclusive
  and differential fiducial production cross sections in the diphoton decay
  channel with pp collisions at $ \sqrt{s} $ = 13 TeV}},
  \href{https://doi.org/10.1007/JHEP07(2023)091}{\emph{JHEP} {\bfseries 07}
  (2023) 091} [\href{https://arxiv.org/abs/2208.12279}{{\ttfamily
  2208.12279}}].

\bibitem{CMS:2022urr}
{\scshape CMS} collaboration, \emph{{Search for the Higgs boson decay to a pair
  of electrons in proton-proton collisions at s=13TeV}},
  \href{https://doi.org/10.1016/j.physletb.2023.137783}{\emph{Phys. Lett. B}
  {\bfseries 846} (2023) 137783}
  [\href{https://arxiv.org/abs/2208.00265}{{\ttfamily 2208.00265}}].

\bibitem{CMS:2022kdx}
{\scshape CMS} collaboration, \emph{{Search for Higgs boson pairs decaying to
  WW*WW*, WW*$\tau\tau$, and $\tau\tau\tau\tau$ in proton-proton collisions at
  $\sqrt{s}$ = 13 TeV}},
  \href{https://doi.org/10.1007/JHEP07(2023)095}{\emph{JHEP} {\bfseries 07}
  (2023) 095} [\href{https://arxiv.org/abs/2206.10268}{{\ttfamily
  2206.10268}}].

\bibitem{CMS:2022uhn}
{\scshape CMS} collaboration, \emph{{Measurements of the Higgs boson production
  cross section and couplings in the W boson pair decay channel in
  proton-proton collisions at $\sqrt{s}=13\,\text {Te\hspace{-.08em}V} $}},
  \href{https://doi.org/10.1140/epjc/s10052-023-11632-6}{\emph{Eur. Phys. J. C}
  {\bfseries 83} (2023) 667}
  [\href{https://arxiv.org/abs/2206.09466}{{\ttfamily 2206.09466}}].

\bibitem{CMS:2022dwd}
{\scshape CMS} collaboration, \emph{{A portrait of the Higgs boson by the CMS
  experiment ten years after the discovery.}},
  \href{https://doi.org/10.1038/s41586-022-04892-x}{\emph{Nature} {\bfseries
  607} (2022) 60} [\href{https://arxiv.org/abs/2207.00043}{{\ttfamily
  2207.00043}}].

\bibitem{Buchmuller:1985jz}
W.~Buchmuller and D.~Wyler, \emph{{Effective Lagrangian Analysis of New
  Interactions and Flavor Conservation}},
  \href{https://doi.org/10.1016/0550-3213(86)90262-2}{\emph{Nucl. Phys. B}
  {\bfseries 268} (1986) 621}.

\bibitem{Azatov:2016sqh}
A.~Azatov, R.~Contino, C.~S. Machado and F.~Riva, \emph{{Helicity selection
  rules and noninterference for BSM amplitudes}},
  \href{https://doi.org/10.1103/PhysRevD.95.065014}{\emph{Phys. Rev. D}
  {\bfseries 95} (2017) 065014}
  [\href{https://arxiv.org/abs/1607.05236}{{\ttfamily 1607.05236}}].

\bibitem{Corbett:2015cfe}
T.~Corbett, \emph{{Effective Lagrangians for Higgs physics}}, Ph.D. thesis,
  Stony Brook U., 4, 2015.

\bibitem{Degrande:2012wf}
C.~Degrande, N.~Greiner, W.~Kilian, O.~Mattelaer, H.~Mebane, T.~Stelzer et~al.,
  \emph{{Effective Field Theory: A Modern Approach to Anomalous Couplings}},
  \href{https://doi.org/10.1016/j.aop.2013.04.016}{\emph{Annals Phys.}
  {\bfseries 335} (2013) 21} [\href{https://arxiv.org/abs/1205.4231}{{\ttfamily
  1205.4231}}].

\bibitem{Degrande:2013rea}
C.~Degrande, O.~Eboli, B.~Feigl, B.~Jager, W.~Kilian, O.~Mattelaer et~al.,
  \emph{{Monte Carlo tools for studies of non-standard electroweak gauge boson
  interactions in multi-boson processes: A Snowmass White Paper}},  in
  \emph{{Snowmass 2013}: {Snowmass on the Mississippi}}, 9, 2013,
  \href{https://arxiv.org/abs/1309.7890}{{\ttfamily 1309.7890}}.

\bibitem{Hagiwara:1993ck}
K.~Hagiwara, S.~Ishihara, R.~Szalapski and D.~Zeppenfeld, \emph{{Low-energy
  effects of new interactions in the electroweak boson sector}},
  \href{https://doi.org/10.1103/PhysRevD.48.2182}{\emph{Phys. Rev. D}
  {\bfseries 48} (1993) 2182}.

\bibitem{Hagiwara:1986vm}
K.~Hagiwara, R.~D. Peccei, D.~Zeppenfeld and K.~Hikasa, \emph{{Probing the Weak
  Boson Sector in e+ e- ---\ensuremath{>} W+ W-}},
  \href{https://doi.org/10.1016/0550-3213(87)90685-7}{\emph{Nucl. Phys. B}
  {\bfseries 282} (1987) 253}.

\bibitem{Subba:2023rpm}
A.~Subba and R.~K. Singh, \emph{{Study of anomalous $W^-W^+\gamma/Z$ couplings
  using polarizations and spin correlations in $e^-e^+\to W^-W^+$ with
  polarized beams}},  \href{https://arxiv.org/abs/2305.15106}{{\ttfamily
  2305.15106}}.

\bibitem{Choudhury:2022iqz}
D.~Choudhury, K.~Deka, S.~Maharana and L.~K. Saini, \emph{{Anomalous gauge
  couplings vis-\`a-vis (g-2)\ensuremath{\mu} and flavor observables}},
  \href{https://doi.org/10.1103/PhysRevD.106.115026}{\emph{Phys. Rev. D}
  {\bfseries 106} (2022) 115026}
  [\href{https://arxiv.org/abs/2203.04673}{{\ttfamily 2203.04673}}].

\bibitem{CMS:2021icx}
{\scshape CMS} collaboration, \emph{{Measurement of the inclusive and
  differential WZ production cross sections, polarization angles, and triple
  gauge couplings in pp collisions at $ \sqrt{s} $ = 13 TeV}},
  \href{https://doi.org/10.1007/JHEP07(2022)032}{\emph{JHEP} {\bfseries 07}
  (2022) 032} [\href{https://arxiv.org/abs/2110.11231}{{\ttfamily
  2110.11231}}].

\bibitem{ATLAS:2021jgw}
{\scshape ATLAS} collaboration, \emph{{Measurements of $W^+W^-+\ge 1~$jet
  production cross-sections in $pp$ collisions at $\sqrt{s}=13~$TeV with the
  ATLAS detector}}, \href{https://doi.org/10.1007/JHEP06(2021)003}{\emph{JHEP}
  {\bfseries 06} (2021) 003}
  [\href{https://arxiv.org/abs/2103.10319}{{\ttfamily 2103.10319}}].

\bibitem{CMS:2021foa}
{\scshape CMS} collaboration, \emph{{Measurement of the W$\gamma$ Production
  Cross Section in Proton-Proton Collisions at $\sqrt {s}$=13\,\,TeV and
  Constraints on Effective Field Theory Coefficients}},
  \href{https://doi.org/10.1103/PhysRevLett.126.252002}{\emph{Phys. Rev. Lett.}
  {\bfseries 126} (2021) 252002}
  [\href{https://arxiv.org/abs/2102.02283}{{\ttfamily 2102.02283}}].

\bibitem{Biekotter:2021int}
A.~Biek\"otter, P.~Gregg, F.~Krauss and M.~Sch\"onherr, \emph{{Constraining CP
  violating operators in charged and neutral triple gauge couplings}},
  \href{https://doi.org/10.1016/j.physletb.2021.136311}{\emph{Phys. Lett. B}
  {\bfseries 817} (2021) 136311}
  [\href{https://arxiv.org/abs/2102.01115}{{\ttfamily 2102.01115}}].

\bibitem{Calfayan:2020tuk}
{\scshape ATLAS} collaboration, \emph{{Measurements of inclusive $WW$ and $WZ$
  production with ATLAS}},
  \href{https://doi.org/10.22323/1.364.0663}{\emph{PoS} {\bfseries EPS-HEP2019}
  (2020) 663}.

\bibitem{Rahaman:2019lab}
R.~Rahaman and R.~K. Singh, \emph{{Unravelling the anomalous gauge boson
  couplings in $ZW^\pm$ production at the LHC and the role of spin-$1$
  polarizations}}, \href{https://doi.org/10.1007/JHEP04(2020)075}{\emph{JHEP}
  {\bfseries 04} (2020) 075}
  [\href{https://arxiv.org/abs/1911.03111}{{\ttfamily 1911.03111}}].

\bibitem{Koksal:2019oqt}
M.~K\"oksal, A.~A. Billur, A.~Guti\'errez-Rodr\'\i{}guez and M.~A.
  Hern\'andez-Ru\'\i{}z, \emph{{Bounds on the non-standard $W^+W^-\gamma$
  couplings at the LHeC and the FCC-he}},
  \href{https://doi.org/10.1016/j.physletb.2020.135661}{\emph{Phys. Lett. B}
  {\bfseries 808} (2020) 135661}
  [\href{https://arxiv.org/abs/1910.06747}{{\ttfamily 1910.06747}}].

\bibitem{Gutierrez-Rodriguez:2019hek}
A.~Guti\'errez-Rodr\'\i{}guez, M.~K\"oksal, A.~A. Billur and M.~A.
  Hern\'andez-Ru\'\i{}z, \emph{{Probing model-independent limits on
  $W^+W^-\gamma$ triple gauge boson vertex at the LHeC and the FCC-he}},
  \href{https://doi.org/10.1088/1361-6471/ab7ff9}{\emph{J. Phys. G} {\bfseries
  47} (2020) 055005} [\href{https://arxiv.org/abs/1910.02307}{{\ttfamily
  1910.02307}}].

\bibitem{Billur:2019cav}
A.~A. Billur, M.~K\"oksal, A.~Guti\'errez-Rodr\'\i{}guez and M.~A.
  Hern\'andez-Ru\'\i{}z, \emph{{Model-independent limits for anomalous triple
  gauge bosons $W^+W^-\gamma $ couplings at the CLIC}},
  \href{https://doi.org/10.1140/epjp/s13360-021-01684-6}{\emph{Eur. Phys. J.
  Plus} {\bfseries 136} (2021) 697}
  [\href{https://arxiv.org/abs/1909.10299}{{\ttfamily 1909.10299}}].

\bibitem{Rahaman:2019mnz}
R.~Rahaman and R.~K. Singh, \emph{{Probing the anomalous triple gauge boson
  couplings in $e^+e^-\to W^+W^-$ using $W$ polarizations with polarized
  beams}}, \href{https://doi.org/10.1103/PhysRevD.101.075044}{\emph{Phys. Rev.
  D} {\bfseries 101} (2020) 075044}
  [\href{https://arxiv.org/abs/1909.05496}{{\ttfamily 1909.05496}}].

\bibitem{CMS:2019ppl}
{\scshape CMS} collaboration, \emph{{Search for anomalous triple gauge
  couplings in WW and WZ production in lepton + jet events in proton-proton
  collisions at $\sqrt{s} =$ 13 TeV}},
  \href{https://doi.org/10.1007/JHEP12(2019)062}{\emph{JHEP} {\bfseries 12}
  (2019) 062} [\href{https://arxiv.org/abs/1907.08354}{{\ttfamily
  1907.08354}}].

\bibitem{CMS:2019efc}
{\scshape CMS} collaboration, \emph{{Measurements of the pp $\to$ WZ inclusive
  and differential production cross section and constraints on charged
  anomalous triple gauge couplings at $\sqrt{s} =$ 13 TeV}},
  \href{https://doi.org/10.1007/JHEP04(2019)122}{\emph{JHEP} {\bfseries 04}
  (2019) 122} [\href{https://arxiv.org/abs/1901.03428}{{\ttfamily
  1901.03428}}].

\bibitem{daSilvaAlmeida:2018iqo}
E.~da~Silva~Almeida, A.~Alves, N.~Rosa~Agostinho, O.~J.~P. \'Eboli and M.~C.
  Gonzalez-Garcia, \emph{{Electroweak Sector Under Scrutiny: A Combined
  Analysis of LHC and Electroweak Precision Data}},
  \href{https://doi.org/10.1103/PhysRevD.99.033001}{\emph{Phys. Rev. D}
  {\bfseries 99} (2019) 033001}
  [\href{https://arxiv.org/abs/1812.01009}{{\ttfamily 1812.01009}}].

\bibitem{Bellan:2018xxs}
{\scshape CMS} collaboration, \emph{{Multiboson production measurements at the
  CMS experiment}}, \href{https://doi.org/10.22323/1.340.0194}{\emph{PoS}
  {\bfseries ICHEP2018} (2019) 194}.

\bibitem{Cuevas:2018jah}
{\scshape CMS} collaboration, \emph{{Latest results on diboson and multiboson
  production from the CMS experiment}},
  \href{https://doi.org/10.22323/1.321.0288}{\emph{PoS} {\bfseries LHCP2018}
  (2018) 288}.

\bibitem{CMS:2018hlo}
{\scshape CMS} collaboration, \emph{{Measurements of the
  $\mathrm{pp}\to\mathrm{WZ}$ inclusive and differential production cross
  section and constraints on charged anomalous triple gauge couplings at
  $\sqrt{s} = 13~\mathrm{TeV}$.}}, .

\bibitem{Li:2017kfk}
R.~Li, X.-M. Shen, K.~Wang, T.~Xu, L.~Zhang and G.~Zhu, \emph{{Probing
  anomalous $WW\gamma$ triple gauge bosons coupling at the LHeC}},
  \href{https://doi.org/10.1103/PhysRevD.97.075043}{\emph{Phys. Rev. D}
  {\bfseries 97} (2018) 075043}
  [\href{https://arxiv.org/abs/1711.05607}{{\ttfamily 1711.05607}}].

\bibitem{Burger:2017goy}
{\scshape ATLAS} collaboration, \emph{{Measurement of the diboson production
  cross section at 8TeV and 13TeV and limits on anomalous triple gauge
  couplings with the ATLAS detector}},
  \href{https://doi.org/10.22323/1.297.0155}{\emph{PoS} {\bfseries DIS2017}
  (2018) 155}.

\bibitem{ATLAS:2017pbb}
{\scshape ATLAS} collaboration, \emph{{Measurement of $WW/WZ \to \ell \nu q
  q^{\prime}$ production with the hadronically decaying boson reconstructed as
  one or two jets in $pp$ collisions at $\sqrt{s}=8$ TeV with ATLAS, and
  constraints on anomalous gauge couplings}},
  \href{https://doi.org/10.1140/epjc/s10052-017-5084-2}{\emph{Eur. Phys. J. C}
  {\bfseries 77} (2017) 563}
  [\href{https://arxiv.org/abs/1706.01702}{{\ttfamily 1706.01702}}].

\bibitem{CMS:2017egm}
{\scshape CMS} collaboration, \emph{{Search for anomalous couplings in boosted
  $\mathrm{ WW/WZ }\to\ell\nu\mathrm{ q \bar{q} }$ production in proton-proton
  collisions at $\sqrt{s} =$ 8 TeV}},
  \href{https://doi.org/10.1016/j.physletb.2017.06.009}{\emph{Phys. Lett. B}
  {\bfseries 772} (2017) 21}
  [\href{https://arxiv.org/abs/1703.06095}{{\ttfamily 1703.06095}}].

\bibitem{ATLAS:2017luz}
{\scshape ATLAS} collaboration, \emph{{Measurements of electroweak $Wjj$
  production and constraints on anomalous gauge couplings with the ATLAS
  detector}}, \href{https://doi.org/10.1140/epjc/s10052-017-5007-2}{\emph{Eur.
  Phys. J. C} {\bfseries 77} (2017) 474}
  [\href{https://arxiv.org/abs/1703.04362}{{\ttfamily 1703.04362}}].

\bibitem{Iliadis:2017bqd}
{\scshape ATLAS} collaboration, \emph{{Measurement of the $WZ$ boson pair
  production cross section at 13 TeV and confidence intervals on anomalous
  triple gauge couplings with the ATLAS detector}},
  \href{https://doi.org/10.1051/epjconf/201713708009}{\emph{EPJ Web Conf.}
  {\bfseries 137} (2017) 08009}.

\bibitem{CMS:2016gct}
{\scshape CMS} collaboration, \emph{{Measurement of electroweak-induced
  production of W$\gamma$ with two jets in pp collisions at $ \sqrt{s}=8 $ TeV
  and constraints on anomalous quartic gauge couplings}},
  \href{https://doi.org/10.1007/JHEP06(2017)106}{\emph{JHEP} {\bfseries 06}
  (2017) 106} [\href{https://arxiv.org/abs/1612.09256}{{\ttfamily
  1612.09256}}].

\bibitem{Hassani:2016jnz}
{\scshape ATLAS} collaboration, \emph{{Measurement of the $W^+W^- \to \ell \nu
  \ell \nu$ production cross section at $\sqrt s = $ 8 TeV and 13 TeV and
  limits on anomalous triple gauge couplings with the ATLAS detector}},
  \href{https://doi.org/10.22323/1.282.0670}{\emph{PoS} {\bfseries ICHEP2016}
  (2016) 670}.

\bibitem{Becker:2016fju}
{\scshape ATLAS} collaboration, \emph{{Measurements of Multi-boson production,
  Trilinear and Quartic Gauge Couplings with the ATLAS detector}},
  \href{https://doi.org/10.1051/epjconf/201612604005}{\emph{EPJ Web Conf.}
  {\bfseries 126} (2016) 04005}.

\bibitem{Wang:2016zbh}
{\scshape ATLAS} collaboration, \emph{{Vector boson scattering, triple
  gauge-boson production and limits on anomalous quartic gauge-boson couplings
  with the ATLAS detector}},
  \href{https://doi.org/10.22323/1.282.0685}{\emph{PoS} {\bfseries ICHEP2016}
  (2016) 685}.

\bibitem{Zhang:2016zsp}
Z.~Zhang, \emph{{Time to Go Beyond Triple-Gauge-Boson-Coupling Interpretation
  of $W$ Pair Production}},
  \href{https://doi.org/10.1103/PhysRevLett.118.011803}{\emph{Phys. Rev. Lett.}
  {\bfseries 118} (2017) 011803}
  [\href{https://arxiv.org/abs/1610.01618}{{\ttfamily 1610.01618}}].

\bibitem{Falkowski:2016cxu}
A.~Falkowski, M.~Gonzalez-Alonso, A.~Greljo, D.~Marzocca and M.~Son,
  \emph{{Anomalous Triple Gauge Couplings in the Effective Field Theory
  Approach at the LHC}},
  \href{https://doi.org/10.1007/JHEP02(2017)115}{\emph{JHEP} {\bfseries 02}
  (2017) 115} [\href{https://arxiv.org/abs/1609.06312}{{\ttfamily
  1609.06312}}].

\bibitem{CMS:2016qth}
{\scshape CMS} collaboration, \emph{{Measurement of the WZ production cross
  section in pp collisions at $\sqrt{s} = 7$ and 8 $\,\text{TeV}$ and search
  for anomalous triple gauge couplings at $\sqrt{s} = 8\,\text{TeV} $}},
  \href{https://doi.org/10.1140/epjc/s10052-017-4730-z}{\emph{Eur. Phys. J. C}
  {\bfseries 77} (2017) 236}
  [\href{https://arxiv.org/abs/1609.05721}{{\ttfamily 1609.05721}}].

\bibitem{ATLAS:2016qzn}
{\scshape ATLAS} collaboration, \emph{{Measurement of $W^{\pm}Z$ boson
  pair-production in $pp$ collisions at $\sqrt{s}=13$ TeV with the ATLAS
  Detector and confidence intervals for anomalous triple gauge boson
  couplings}}, .

\bibitem{Etesami:2016rwu}
S.~M. Etesami, S.~Khatibi and M.~Mohammadi~Najafabadi, \emph{{Measuring
  anomalous WW $\gamma $ and t $\bar{\text {t}}\gamma $ couplings using top+
  $\gamma $ production at the LHC}},
  \href{https://doi.org/10.1140/epjc/s10052-016-4376-2}{\emph{Eur. Phys. J. C}
  {\bfseries 76} (2016) 533}
  [\href{https://arxiv.org/abs/1606.02178}{{\ttfamily 1606.02178}}].

\bibitem{Falkowski:2015jaa}
A.~Falkowski, M.~Gonzalez-Alonso, A.~Greljo and D.~Marzocca, \emph{{Global
  constraints on anomalous triple gauge couplings in effective field theory
  approach}}, \href{https://doi.org/10.1103/PhysRevLett.116.011801}{\emph{Phys.
  Rev. Lett.} {\bfseries 116} (2016) 011801}
  [\href{https://arxiv.org/abs/1508.00581}{{\ttfamily 1508.00581}}].

\bibitem{ATLAS:2014ofc}
{\scshape ATLAS} collaboration, \emph{{Measurement of the $WW+WZ$ cross section
  and limits on anomalous triple gauge couplings using final states with one
  lepton, missing transverse momentum, and two jets with the ATLAS detector at
  $\sqrt{\rm{s}} = 7$ TeV}},
  \href{https://doi.org/10.1007/JHEP01(2015)049}{\emph{JHEP} {\bfseries 01}
  (2015) 049} [\href{https://arxiv.org/abs/1410.7238}{{\ttfamily 1410.7238}}].

\bibitem{Biswal:2014oaa}
S.~S. Biswal, M.~Patra and S.~Raychaudhuri, \emph{{Anomalous Triple Gauge
  Vertices at the Large Hadron-Electron Collider}},
  \href{https://arxiv.org/abs/1405.6056}{{\ttfamily 1405.6056}}.

\bibitem{CMS:2013ryd}
{\scshape CMS} collaboration, \emph{{Measurement of the $W\gamma$ and $Z\gamma$
  Inclusive Cross Sections in $pp$ Collisions at $\sqrt s=7$ TeV and Limits on
  Anomalous Triple Gauge Boson Couplings}},
  \href{https://doi.org/10.1103/PhysRevD.89.092005}{\emph{Phys. Rev. D}
  {\bfseries 89} (2014) 092005}
  [\href{https://arxiv.org/abs/1308.6832}{{\ttfamily 1308.6832}}].

\bibitem{Han:2013jmw}
{\scshape ATLAS} collaboration, \emph{{Measurement of W \ensuremath{\gamma} and
  Z \ensuremath{\gamma} production cross sections in pp collisions at 7 TeV and
  limits on anomalous triple gauge couplings with the ATLAS detector}},
  \href{https://doi.org/10.22323/1.174.0071}{\emph{PoS} {\bfseries ICHEP2012}
  (2013) 071}.

\bibitem{Corbett:2012ja}
T.~Corbett, O.~J.~P. Eboli, J.~Gonzalez-Fraile and M.~C. Gonzalez-Garcia,
  \emph{{Robust Determination of the Higgs Couplings: Power to the Data}},
  \href{https://doi.org/10.1103/PhysRevD.87.015022}{\emph{Phys. Rev. D}
  {\bfseries 87} (2013) 015022}
  [\href{https://arxiv.org/abs/1211.4580}{{\ttfamily 1211.4580}}].

\bibitem{Hernandez-Juarez:2024zpk}
A.~I. Hern\'andez-Ju\'arez, R.~Gait\'an and G.~Tavares-Velasco,
  \emph{{Polarized and unpolarized off-shell $H^\ast\to ZZ\rightarrow 4\ell$
  decay above the $2m_Z$ threshold}},
  \href{https://arxiv.org/abs/2402.18497}{{\ttfamily 2402.18497}}.

\bibitem{Kniehl:1990mq}
B.~A. Kniehl, \emph{{Radiative corrections for $H \to Z Z$ in the standard
  model}}, \href{https://doi.org/10.1016/0550-3213(91)90126-I}{\emph{Nucl.
  Phys. B} {\bfseries 352} (1991) 1}.

\bibitem{Soni:1993jc}
A.~Soni and R.~M. Xu, \emph{{Probing CP violation via Higgs decays to four
  leptons}}, \href{https://doi.org/10.1103/PhysRevD.48.5259}{\emph{Phys. Rev.
  D} {\bfseries 48} (1993) 5259}
  [\href{https://arxiv.org/abs/hep-ph/9301225}{{\ttfamily hep-ph/9301225}}].

\bibitem{Hernandez-Juarez:2023dor}
A.~I. Hern\'andez-Ju\'arez, G.~Tavares-Velasco and A.~Fern\'andez-T\'ellez,
  \emph{{New evaluation of the HZZ coupling: Direct bounds on anomalous
  contributions and CP-violating effects via a new asymmetry}},
  \href{https://doi.org/10.1103/PhysRevD.107.115031}{\emph{Phys. Rev. D}
  {\bfseries 107} (2023) 115031}
  [\href{https://arxiv.org/abs/2301.13127}{{\ttfamily 2301.13127}}].

\bibitem{Biswal:2005fh}
S.~S. Biswal, R.~M. Godbole, R.~K. Singh and D.~Choudhury, \emph{{Signatures of
  anomalous VVH interactions at a linear collider}},
  \href{https://doi.org/10.1103/PhysRevD.74.039904}{\emph{Phys. Rev. D}
  {\bfseries 73} (2006) 035001}
  [\href{https://arxiv.org/abs/hep-ph/0509070}{{\ttfamily hep-ph/0509070}}].

\bibitem{Dutta:2008bh}
S.~Dutta, K.~Hagiwara and Y.~Matsumoto, \emph{{Measuring the Higgs-Vector boson
  Couplings at Linear $e^{+} e^{-}$ Collider}},
  \href{https://doi.org/10.1103/PhysRevD.78.115016}{\emph{Phys. Rev. D}
  {\bfseries 78} (2008) 115016}
  [\href{https://arxiv.org/abs/0808.0477}{{\ttfamily 0808.0477}}].

\bibitem{Rao:2020hel}
K.~Rao, S.~D. Rindani and P.~Sarmah, \emph{{Study of anomalous gauge-Higgs
  couplings using Z boson polarization at LHC}},
  \href{https://doi.org/10.1016/j.nuclphysb.2021.115317}{\emph{Nucl. Phys. B}
  {\bfseries 964} (2021) 115317}
  [\href{https://arxiv.org/abs/2009.00980}{{\ttfamily 2009.00980}}].

\bibitem{Godbole:2007cn}
R.~M. Godbole, D.~J. Miller and M.~M. Muhlleitner, \emph{{Aspects of CP
  violation in the H ZZ coupling at the LHC}},
  \href{https://doi.org/10.1088/1126-6708/2007/12/031}{\emph{JHEP} {\bfseries
  12} (2007) 031} [\href{https://arxiv.org/abs/0708.0458}{{\ttfamily
  0708.0458}}].

\bibitem{Cakir:2013bxa}
I.~T. Cakir, O.~Cakir, A.~Senol and A.~T. Tasci, \emph{{Probing Anomalous HZZ
  Couplings at the LHeC}},
  \href{https://doi.org/10.1142/S0217732313501423}{\emph{Mod. Phys. Lett. A}
  {\bfseries 28} (2013) 1350142}
  [\href{https://arxiv.org/abs/1304.3616}{{\ttfamily 1304.3616}}].

\bibitem{Sahin:2019wew}
B.~\c{S}ahin, \emph{{Search for the anomalous ZZH couplings at the CLIC}},
  \href{https://doi.org/10.1142/S0217732319502997}{\emph{Mod. Phys. Lett. A}
  {\bfseries 34} (2019) 1950299}.

\bibitem{Gauld:2023gtb}
R.~Gauld, U.~Haisch and L.~Schnell, \emph{{SMEFT at NNLO+PS: Vh production}},
  \href{https://doi.org/10.1007/JHEP01(2024)192}{\emph{JHEP} {\bfseries 01}
  (2024) 192} [\href{https://arxiv.org/abs/2311.06107}{{\ttfamily
  2311.06107}}].

\bibitem{Gounaris:2000tb}
G.~J. Gounaris, J.~Layssac and F.~M. Renard, \emph{{New and standard physics
  contributions to anomalous Z and gamma selfcouplings}},
  \href{https://doi.org/10.1103/PhysRevD.62.073013}{\emph{Phys. Rev. D}
  {\bfseries 62} (2000) 073013}
  [\href{https://arxiv.org/abs/hep-ph/0003143}{{\ttfamily hep-ph/0003143}}].

\bibitem{He:2019kgh}
H.-R. He, X.~Wan and Y.-K. Wang, \emph{{Anomalous $H\to ZZ \to 4\ell$ decay and
  its interference effects on gluon\textendash{}gluon contribution at the
  LHC}}, \href{https://doi.org/10.1088/1674-1137/abb4c8}{\emph{Chin. Phys. C}
  {\bfseries 44} (2020) 123101}
  [\href{https://arxiv.org/abs/1902.04756}{{\ttfamily 1902.04756}}].

\bibitem{Buchalla:2013mpa}
G.~Buchalla, O.~Cata and G.~D'Ambrosio, \emph{{Nonstandard Higgs couplings from
  angular distributions in $h\to Z \ell^+\ell^-$}},
  \href{https://doi.org/10.1140/epjc/s10052-014-2798-2}{\emph{Eur. Phys. J. C}
  {\bfseries 74} (2014) 2798}
  [\href{https://arxiv.org/abs/1310.2574}{{\ttfamily 1310.2574}}].

\bibitem{Lagouri:2024ozw}
T.~Lagouri, \emph{{Measurements of Higgs boson couplings and simplified
  template cross sections in bosonic final states (WW*, ZZ*,
  \ensuremath{\gamma}\ensuremath{\gamma}) at the ATLAS experiment}},
  \href{https://doi.org/10.22323/1.449.0381}{\emph{PoS} {\bfseries EPS-HEP2023}
  (2024) 381}.

\bibitem{Spor:2023sdk}
S.~Spor, \emph{{Study on the sensitivity of the Higgs boson couplings in
  photon-photon collision at CLIC and muon collider}},
  \href{https://arxiv.org/abs/2309.12498}{{\ttfamily 2309.12498}}.

\bibitem{DAgnolo:2023rnh}
R.~T. D'Agnolo, F.~Nortier, G.~Rigo and P.~Sesma, \emph{{The two scales of new
  physics in Higgs couplings}},
  \href{https://doi.org/10.1007/JHEP08(2023)019}{\emph{JHEP} {\bfseries 08}
  (2023) 019} [\href{https://arxiv.org/abs/2305.19325}{{\ttfamily
  2305.19325}}].

\bibitem{Fabbrichesi:2023jep}
M.~Fabbrichesi, R.~Floreanini, E.~Gabrielli and L.~Marzola, \emph{{Stringent
  bounds on HWW and HZZ anomalous couplings with quantum tomography at the
  LHC}}, \href{https://doi.org/10.1007/JHEP09(2023)195}{\emph{JHEP} {\bfseries
  09} (2023) 195} [\href{https://arxiv.org/abs/2304.02403}{{\ttfamily
  2304.02403}}].

\bibitem{Lu:2022jzs}
Y.-S. Lu, Y.-K. Wang and X.-Y. You, \emph{{Study of $HZZ$ anomalous couplings
  by angular differential cross sections}},
  \href{https://arxiv.org/abs/2211.07478}{{\ttfamily 2211.07478}}.

\bibitem{Sharma:2022epc}
P.~Sharma and A.~Shivaji, \emph{{Probing non-standard HVV (V = W, Z) couplings
  in single Higgs production at future electron-proton collider}},
  \href{https://doi.org/10.1007/JHEP10(2022)108}{\emph{JHEP} {\bfseries 10}
  (2022) 108} [\href{https://arxiv.org/abs/2207.03862}{{\ttfamily
  2207.03862}}].

\bibitem{Kumar:2019bmk}
S.~Kumar, P.~Poulose, R.~Rahaman and R.~K. Singh, \emph{{Measuring Higgs
  self-couplings in the presence of VVH and VVHH at the ILC}},
  \href{https://doi.org/10.1142/S0217751X19500945}{\emph{Int. J. Mod. Phys. A}
  {\bfseries 34} (2019) 1950094}
  [\href{https://arxiv.org/abs/1905.06601}{{\ttfamily 1905.06601}}].

\bibitem{Gabrielli:2013era}
E.~Gabrielli, M.~Heikinheimo, L.~Marzola, B.~Mele, C.~Spethmann and H.~Veermae,
  \emph{{Anomalous Higgs-boson coupling effects in HW+W\ensuremath{-}
  production at the LHC}},
  \href{https://doi.org/10.1103/PhysRevD.89.053012}{\emph{Phys. Rev. D}
  {\bfseries 89} (2014) 053012}
  [\href{https://arxiv.org/abs/1312.4956}{{\ttfamily 1312.4956}}].

\bibitem{Gonzalez-Lopez:2020lpd}
M.~Gonzalez-Lopez, M.~J. Herrero and P.~Martinez-Suarez, \emph{{Testing
  anomalous $H-W$ couplings and Higgs self-couplings via double and triple
  Higgs production at $e^+e^-$ colliders}},
  \href{https://doi.org/10.1140/epjc/s10052-021-09048-1}{\emph{Eur. Phys. J. C}
  {\bfseries 81} (2021) 260}
  [\href{https://arxiv.org/abs/2011.13915}{{\ttfamily 2011.13915}}].

\bibitem{Asteriadis:2022ebf}
K.~Asteriadis, F.~Caola, K.~Melnikov and R.~R\"ontsch, \emph{{Anomalous Higgs
  boson couplings in weak boson fusion production at NNLO in QCD}},
  \href{https://doi.org/10.1103/PhysRevD.107.034034}{\emph{Phys. Rev. D}
  {\bfseries 107} (2023) 034034}
  [\href{https://arxiv.org/abs/2206.14630}{{\ttfamily 2206.14630}}].

\bibitem{Senol:2012fc}
A.~Senol, \emph{{Anomalous Higgs Couplings at the LHeC}},
  \href{https://doi.org/10.1016/j.nuclphysb.2013.04.016}{\emph{Nucl. Phys. B}
  {\bfseries 873} (2013) 293}
  [\href{https://arxiv.org/abs/1212.6869}{{\ttfamily 1212.6869}}].

\bibitem{Buchalla:2015wfa}
G.~Buchalla, O.~Cata, A.~Celis and C.~Krause, \emph{{Note on Anomalous
  Higgs-Boson Couplings in Effective Field Theory}},
  \href{https://doi.org/10.1016/j.physletb.2015.09.027}{\emph{Phys. Lett. B}
  {\bfseries 750} (2015) 298}
  [\href{https://arxiv.org/abs/1504.01707}{{\ttfamily 1504.01707}}].

\bibitem{Biswal:2012mp}
S.~S. Biswal, R.~M. Godbole, B.~Mellado and S.~Raychaudhuri, \emph{{Azimuthal
  Angle Probe of Anomalous $HWW$ Couplings at a High Energy $ep$ Collider}},
  \href{https://doi.org/10.1103/PhysRevLett.109.261801}{\emph{Phys. Rev. Lett.}
  {\bfseries 109} (2012) 261801}
  [\href{https://arxiv.org/abs/1203.6285}{{\ttfamily 1203.6285}}].

\bibitem{Corbett:2012dm}
T.~Corbett, O.~J.~P. Eboli, J.~Gonzalez-Fraile and M.~C. Gonzalez-Garcia,
  \emph{{Constraining anomalous Higgs interactions}},
  \href{https://doi.org/10.1103/PhysRevD.86.075013}{\emph{Phys. Rev. D}
  {\bfseries 86} (2012) 075013}
  [\href{https://arxiv.org/abs/1207.1344}{{\ttfamily 1207.1344}}].

\bibitem{Davis:2021tiv}
J.~Davis, A.~V. Gritsan, L.~S.~M. Guerra, S.~Kyriacou, J.~Roskes and
  M.~Schulze, \emph{{Constraining anomalous Higgs boson couplings to virtual
  photons}}, \href{https://doi.org/10.1103/PhysRevD.105.096027}{\emph{Phys.
  Rev. D} {\bfseries 105} (2022) 096027}
  [\href{https://arxiv.org/abs/2109.13363}{{\ttfamily 2109.13363}}].

\bibitem{Shi:2018lqf}
L.~Shi, Z.~Liang, B.~Liu and Z.~He, \emph{{Constraining the anomalous Higgs
  boson coupling in $H$+$\gamma$ production}},
  \href{https://doi.org/10.1088/1674-1137/43/4/043001}{\emph{Chin. Phys. C}
  {\bfseries 43} (2019) 043001}
  [\href{https://arxiv.org/abs/1811.02261}{{\ttfamily 1811.02261}}].

\bibitem{L3:2004vpt}
{\scshape L3} collaboration, \emph{{Search for anomalous couplings in the Higgs
  sector at LEP}},
  \href{https://doi.org/10.1016/j.physletb.2004.03.048}{\emph{Phys. Lett. B}
  {\bfseries 589} (2004) 89}
  [\href{https://arxiv.org/abs/hep-ex/0403037}{{\ttfamily hep-ex/0403037}}].

\bibitem{Gonzalez-Garcia:1999ije}
M.~C. Gonzalez-Garcia, \emph{{Anomalous Higgs couplings}},
  \href{https://doi.org/10.1142/S0217751X99001494}{\emph{Int. J. Mod. Phys. A}
  {\bfseries 14} (1999) 3121}
  [\href{https://arxiv.org/abs/hep-ph/9902321}{{\ttfamily hep-ph/9902321}}].

\bibitem{Simmons:1989zs}
E.~H. Simmons, \emph{{Dimension-six Gluon Operators as Probes of New Physics}},
  \href{https://doi.org/10.1016/0370-2693(89)90301-8}{\emph{Phys. Lett. B}
  {\bfseries 226} (1989) 132}.

\bibitem{Dixon:1993xd}
L.~J. Dixon and Y.~Shadmi, \emph{{Testing gluon selfinteractions in three jet
  events at hadron colliders}},
  \href{https://doi.org/10.1016/0550-3213(94)90563-0}{\emph{Nucl. Phys. B}
  {\bfseries 423} (1994) 3}
  [\href{https://arxiv.org/abs/hep-ph/9312363}{{\ttfamily hep-ph/9312363}}].

\bibitem{Alloul:2013bka}
A.~Alloul, N.~D. Christensen, C.~Degrande, C.~Duhr and B.~Fuks,
  \emph{{FeynRules 2.0 - A complete toolbox for tree-level phenomenology}},
  \href{https://doi.org/10.1016/j.cpc.2014.04.012}{\emph{Comput. Phys. Commun.}
  {\bfseries 185} (2014) 2250}
  [\href{https://arxiv.org/abs/1310.1921}{{\ttfamily 1310.1921}}].

\bibitem{Darme:2023jdn}
L.~Darm\'e et~al., \emph{{UFO 2.0: the \textquoteleft{}Universal Feynman
  Output\textquoteright{} format}},
  \href{https://doi.org/10.1140/epjc/s10052-023-11780-9}{\emph{Eur. Phys. J. C}
  {\bfseries 83} (2023) 631}
  [\href{https://arxiv.org/abs/2304.09883}{{\ttfamily 2304.09883}}].

\bibitem{Degrande:2011ua}
C.~Degrande, C.~Duhr, B.~Fuks, D.~Grellscheid, O.~Mattelaer and T.~Reiter,
  \emph{{UFO - The Universal FeynRules Output}},
  \href{https://doi.org/10.1016/j.cpc.2012.01.022}{\emph{Comput. Phys. Commun.}
  {\bfseries 183} (2012) 1201}
  [\href{https://arxiv.org/abs/1108.2040}{{\ttfamily 1108.2040}}].

\bibitem{Alwall:2014hca}
J.~Alwall, R.~Frederix, S.~Frixione, V.~Hirschi, F.~Maltoni, O.~Mattelaer
  et~al., \emph{{The automated computation of tree-level and next-to-leading
  order differential cross sections, and their matching to parton shower
  simulations}}, \href{https://doi.org/10.1007/JHEP07(2014)079}{\emph{JHEP}
  {\bfseries 07} (2014) 079} [\href{https://arxiv.org/abs/1405.0301}{{\ttfamily
  1405.0301}}].

\bibitem{FCC:2018evy}
{\scshape FCC} collaboration, \emph{{FCC-ee: The Lepton Collider}: {Future
  Circular Collider Conceptual Design Report Volume 2}},
  \href{https://doi.org/10.1140/epjst/e2019-900045-4}{\emph{Eur. Phys. J. ST}
  {\bfseries 228} (2019) 261}.

\bibitem{CLICdp:2018cto}
{\scshape CLICdp, CLIC} collaboration, \emph{{The Compact Linear Collider
  (CLIC) - 2018 Summary Report}},
  \href{https://arxiv.org/abs/1812.06018}{{\ttfamily 1812.06018}}.

\bibitem{ILC:2007bjz}
{\scshape ILC} collaboration, \emph{{International Linear Collider Reference
  Design Report Volume 2: Physics at the ILC}},
  \href{https://arxiv.org/abs/0709.1893}{{\ttfamily 0709.1893}}.

\bibitem{ILC:2007oiw}
{\scshape ILC} collaboration, \emph{{ILC Reference Design Report Volume 1 -
  Executive Summary}},  \href{https://arxiv.org/abs/0712.1950}{{\ttfamily
  0712.1950}}.

\bibitem{Moortgat-Pick:2005jsx}
G.~Moortgat-Pick et~al., \emph{{The Role of polarized positrons and electrons
  in revealing fundamental interactions at the linear collider}},
  \href{https://doi.org/10.1016/j.physrep.2007.12.003}{\emph{Phys. Rept.}
  {\bfseries 460} (2008) 131}
  [\href{https://arxiv.org/abs/hep-ph/0507011}{{\ttfamily hep-ph/0507011}}].

\bibitem{Rahaman:2018ujg}
R.~Rahaman and R.~K. Singh, \emph{{Anomalous triple gauge boson couplings in
  $ZZ$ production at the LHC and the role of $Z$ boson polarizations}},
  \href{https://doi.org/10.1016/j.nuclphysb.2019.114754}{\emph{Nucl. Phys. B}
  {\bfseries 948} (2019) 114754}
  [\href{https://arxiv.org/abs/1810.11657}{{\ttfamily 1810.11657}}].

\bibitem{Lewis:2019xzd}
A.~Lewis, \emph{{GetDist: a Python package for analysing Monte Carlo samples}},
   \href{https://arxiv.org/abs/1910.13970}{{\ttfamily 1910.13970}}.

\bibitem{Metropolis:1953am}
N.~Metropolis, A.~W. Rosenbluth, M.~N. Rosenbluth, A.~H. Teller and E.~Teller,
  \emph{{Equation of state calculations by fast computing machines}},
  \href{https://doi.org/10.1063/1.1699114}{\emph{J. Chem. Phys.} {\bfseries 21}
  (1953) 1087}.

\bibitem{Hastings:1970aa}
W.~K. Hastings, \emph{{Monte Carlo Sampling Methods Using Markov Chains and
  Their Applications}},
  \href{https://doi.org/10.1093/biomet/57.1.97}{\emph{Biometrika} {\bfseries
  57} (1970) 97}.

\end{thebibliography}\endgroup
\end{document}